\documentclass[12pt,preprint]{aastex} 

\usepackage{epsf}
\usepackage{enumerate}
\usepackage{amsmath}

\def\nbody{$n$-body}
\def\deg{\ifmmode {^\circ}\else {$^\circ$}\fi}
\def\degree{\ifmmode {^\circ}\else {$^\circ$}\fi}
\def\mum{\ifmmode {\rm \,\mu {\rm m}}\else $\rm \,\mu {\rm m}$\fi}
\def\arcsec{\ifmmode ^{\prime \prime}\else $^{\prime \prime}$\fi}

\def\inch{\ifmmode ^{\prime \prime}\else $^{\prime \prime}$\fi}
\def\Msolar{\ifmmode M_{\odot}\else $ M_{\odot}$\fi}
\def\Mearth{\ifmmode M_{\oplus}\else $ M_{\oplus}$\fi}
\def\Msun{\ifmmode M_{\odot}\else $ M_{\odot}$\fi}
\def\Mjup{\ifmmode M_{J}\else $ M_{J}$\fi}

\def\myC{{\tilde{C}}}
\def\myD{{\tilde{D}}}

\def\epumprate{{\dot{e}_{{\rm pump}}}}
\def\openrate{{\dot{a}_{{\rm in}}}}
\def\ain{a_{{\rm in}}}
\def\aout{a_{{\rm out}}}
\def\tauedamp{\tau_{\rm damp}}
\def\tauadrift{\tau_{\rm drift}}

\def\msunyr{\ifmmode {M_{\odot}~{\rm yr^{-1}}}\else $M_{\odot}~{\rm yr^{-1}}$\fi}
\def\msun{\ifmmode {M_{\odot}}\else $M_{\odot}$\fi}
\def\rsun{\ifmmode {R_{\odot}}\else $R_{\odot}$\fi}
\def\lsun{\ifmmode {L_{\odot}}\else $L_{\odot}$\fi}
\def\mstar{\ifmmode {M_{\star}}\else $M_{\star}$\fi}
\def\rstar{\ifmmode {R_{\star}}\else $R_{\star}$\fi}
\def\tstar{\ifmmode {T_{\star}}\else $T_{\star}$\fi}
\def\lstar{\ifmmode {L_{\star}}\else $L_{\star}$\fi}
\def\abin{{a_{\rm bin}}}
\def\ebin{e_{\rm bin}}
\def\psibin{\psi_{\rm bin}}
\def\varpibin{\varpi_{\rm bin}}

\def\ejup{{e_{\rm J}}}
\def\ajup{{a_{\rm J}}}
\def\omjup{{\Omega_{\rm J}}}

\def\md{\ifmmode {M_d}\else $M_d$\fi}
\def\ld{\ifmmode {L_d}\else $L_d$\fi}
\def\ad{\ifmmode A_d\else $A_d$\fi}
\def\ldlstar{\ifmmode L_d / L_\star\else $L_d / L_{\star}$\fi}
\def\rearth{\ifmmode {\rm R_{\oplus}}\else $\rm R_{\oplus}$\fi}
\def\mearth{\ifmmode {\rm M_{\oplus}}\else $\rm M_{\oplus}$\fi}
\def\qdstar{\ifmmode Q_D^\star\else $Q_D^\star$\fi}
\def\kms{\ifmmode {\rm km~s^{-1}}\else $\rm km~s^{-1}$\fi}
\def\ms{\ifmmode {\rm m~s^{-1}}\else $\rm m~s^{-1}$\fi}
\def\mesc{\ifmmode m_{esc}\else $m_{esc}$\fi}
\def\rmin{\ifmmode r_{min}\else $r_{min}$\fi}
\def\rmax{\ifmmode r_{max}\else $r_{max}$\fi}
\def\mmin{\ifmmode m_{min}\else $m_{min}$\fi}
\def\mmax{\ifmmode m_{max}\else $m_{max}$\fi}
\def\rmind{\ifmmode r_{min,d}\else $r_{min,d}$\fi}
\def\rmaxd{\ifmmode r_{max,d}\else $r_{max,d}$\fi}
\def\mmaxd{\ifmmode m_{max,d}\else $m_{max,d}$\fi}
\def\vrad{\ifmmode v_{rad}\else $v_{rad}$\fi}
\def\rhop{\ifmmode \rho_{{\rm p}}\else $\rho_{{\rm p}}$\fi}

\def\qz{\ifmmode q_{0}\else $q_{0}$\fi}
\def\qi{\ifmmode q_{i}\else $q_{i}$\fi}
\def\ql{\ifmmode q_{l}\else $q_{l}$\fi}
\def\qs{\ifmmode q_{s}\else $q_{s}$\fi}
\def\rbrk{\ifmmode r_{brk}\else $r_{brk}$\fi}
\def\rdamp{\ifmmode r_{damp}\else $r_{damp}$\fi}
\def\rin{\ifmmode r_{in}\else $r_{in}$\fi}
\def\rout{\ifmmode r_{out}\else $r_{out}$\fi}
\def\tin{\ifmmode t_{in}\else $t_{in}$\fi}
\def\tout{\ifmmode t_{out}\else $t_{out}$\fi}
\def\ain{\ifmmode a_{in}\else $a_{in}$\fi}
\def\aout{\ifmmode a_{out}\else $a_{out}$\fi}
\def\r0{\ifmmode R_{0}\else $R_{0}$\fi}
\def\m0{\ifmmode m_{0}\else $m_{0}$\fi}
\def\M0{\ifmmode M_{0}\else $M_{0}$\fi}
\def\Mp{M_1}
\def\Ms{M_2}
\def\Rp{R_1}
\def\Rs{R_2}
\def\phis{\phi_2}

\def\xm{\ifmmode x_{m}\else $x_{m}$\fi}
\def\sigz{\ifmmode \Sigma_0\else $\Sigma_0$\fi}
\def\gyr{\ifmmode {\rm g~yr^{-1}}\else ${\rm g~yr^{-1}}$\fi}
\def\cms{\ifmmode {\rm cm~s^{-1}}\else ${\rm cm~s^{-1}}$\fi}
\def\gcms{\ifmmode {\rm g~cm^{-2}}\else $\rm g~cm^{-2}$\fi}
\def\gcmc{\ifmmode {\rm g~cm^{-3}}\else $\rm g~cm^{-3}$\fi}
\def\atil{\ifmmode {\tilde{a}}\else $\tilde{a}$\fi}
\def\ttil{\ifmmode {\tilde{t}}\else $\tilde{t}$\fi}
\def\sqrttt{\ifmmode {\tilde{t}^{1/2}}\else $\tilde{t}^{1/2}$\fi}

\def\orch{{\it Orchestra}}
\def\orchestra{{\it Orchestra}}

\def\ompresJ{\dot{\varpi}_{\rm p,J}}
\def\ompresS{\dot{\varpi}_{\rm p,S}}
\def\ompreJS{\dot{\varpi}_{\rm J,S}}
\def\ompreJd{\dot{\varpi}_{\rm J,d}}
\def\ompresd{\dot{\varpi}_{\rm p,d}}

\def\ompres{\dot{\varpi}_{\rm p}}
\def\ompreJ{\dot{\varpi}_{\rm J}}

\def\Ravg{\left<{\cal R}\right>}

\def\ompredisk{\dot{\varpi}_{\rm disk}}

\def\eforce{e_{\rm force}}
\def\efree{e_{\rm free}}
\def\nkep{{\Omega_{\rm kep}}}
\def\nbin{{\Omega_{\rm bin}}}

\def\rgc{{R_{\rm g}}}
\def\psigc{{\psi_{\rm g}}}
\def\pagcb{{\gamma}}
\def\ngc{{\Omega_{\rm g}}}

\def\dn{\omega_{\rm rel}}
\def\kappae{\kappa_{\rm e}}
\def\kappai{\kappa_{\rm i}}

\def\SatXXcgs{\left[\frac{\Sigma(a,t)}{\textrm{\small 20\,g/cm$^2$}}\right]}
\def\SinCCcgs{\left[\frac{\Sigma(\ain)}{\textrm{\small 200\,g/cm$^2$}}\right]}
\def\SOhMMcgs{\left[\frac{\Sigma_0}{\textrm{\small 2000\,g/cm$^2$}}\right]}
\def\ejupNow{\left[\frac{\ejup}{\textrm{\small 0.049}}\right]}

\def\aAU{\left[\frac{a}{\textrm{\small 1\,AU}}\right]}
\def\ainXAU{\left[\frac{\ain}{\textrm{\small 10\,AU}}\right]}

\def\mOpIMe{\left[\frac{m}{\textrm{\small 0.1\,M$_\oplus$}}\right]}
\def\hOpOIV{\left[\frac{h_0}{\textrm{\small 0.04\,AU}}\right]}

\def\alphaOpOOOI{\left[\frac{\alpha}{\textrm{\small 1$\times$10$^{-4}$}}\right]}

\def\mIMjup{\left[\frac{m}{\textrm{\small 1\,\Mjup}}\right]}

\def\dv{\Delta v}
\def\vkep{v_{\rm kep}}
\def\vgas{v_{\rm gas}}

\def\tstop{t_{\rm stop}}
\def\rp{r_{{\rm p}}}
\def\mp{m_{{\rm p}}}

\def\ngas{n_{{\rm gas}}}

\def\cs{c_s}

\def\aaero{a_{{\rm aero}}}
\def\ma{\mu}
\def\mfp{\lambda_{\rm mfp}}

\def\circpri{intrabinary}
\def\circpriintro{intrabinary (circumprimary)}

\begin{document}

\title{Terrestrial planet formation: Dynamical shake-up and the 
low mass of Mars}

\author{Benjamin C. Bromley}
\affil{Department of Physics \& Astronomy, University of Utah, 
\\ 115 S 1400 E, Rm 201, Salt Lake City, UT 84112}
\email{bromley@physics.utah.edu}

\author{Scott J. Kenyon}
\affil{Smithsonian Astrophysical Observatory,
\\ 60 Garden St., Cambridge, MA 02138}
\email{skenyon@cfa.harvard.edu}

\begin{abstract}

We consider a dynamical shake-up model to explain the low mass of Mars
and the lack of planets in the asteroid belt. In our scenario, a
secular resonance with Jupiter sweeps through the inner solar system
as the solar nebula depletes, pitting resonant excitation against
collisional damping in the Sun's protoplanetary disk.  We report the
outcome of extensive numerical calculations of planet formation from
planetesimals in the terrestrial zone, with and without dynamical
shake-up. If the Sun's gas disk within the terrestrial zone depletes
in roughly a million years, then the sweeping resonance inhibits
planet formation in the asteroid belt and substantially limits the
size of Mars. This phenomenon likely occurs around other stars with
long-period massive planets, suggesting that asteroid belt analogs are
common.

\end{abstract}


\section{Introduction}

The Sun's rocky planets arose from many small dust particles that were
concentrated into a few larger ones \citep{saf1969}. Although this
process of coagulation, accretion and merging is slow, it is efficient
\citep[e.g.,][]{chambers1998, kb2006}, creating Venus and Earth out of
the primordial dust in the inner solar nebula. Yet just beyond 1~AU,
Mars grew to only a tenth of an Earth mass. The asteroid belt, with
just a few percent of a {Lunar mass} in total, has no planets at
all. Some aspect of planet formation, either its efficiency or the
abundance of solids, prevented the growth of Earth-mass planets beyond
1~AU.  To learn the history of our solar system and to predict the
prevalence of rocky planets throughout the Universe, it is important
to understand the physical processes responsible for a low-mass Mars.

If the Sun's protoplanetary disk was truncated just past the Earth's
orbit, the low mass of Mars and the depletion of solids in the
asteroid belt are natural outcomes \citep{jin2008, hansen2009,
  izidoro2014, walsh2016, haghighipour2016}.  Possibly, the disk was
born that way. However, dynamical excitation and depletion can also
explain the orbital architecture of the inner solar system
\citep{izidoro2015}.  Long-range interactions with Jupiter and Saturn
may provide these excitations \citep{weth1992, naga2005, raymond2009}.
Alternatively, the gas giants themselves might have migrated through
the disk, drifting inward to the terrestrial zone and then outward in
a ``Grand Tack'' that cleared material in their path
\citep{walsh2011}.

After the formation of the gas giants, the remaining gas disk
continues to dissipate from viscous diffusion, photoevaporation, and
erosion by a stellar wind \citep[e.g.,][]{cha2009, alex2009,
  matsuyama2009}.  In the late stages of depletion, the gravity of the
disk and the gravity of the gas giants conspire to generate secular
resonances in the inner solar system \citep{heppenheimer:1980,
  ward1981, lecar:1997, naga2000}.  Within the $\nu_5$ resonance,
where the local apsidal precession rate matches Jupiter's rate of
precession, orbiting bodies experience repeated eccentricity
kicks. As the disk's gravity fades, the location of this resonance
sweeps inward, pumping the eccentricity of all objects in its path
\citep{naga2005, thommes2008}. This dynamical ``shake-up'' sculpts the
inner solar system, leaving its imprint on Mars and the asteroid belt
\citep{obrien2007, nagasawa2007, haghighipour2016}.

Timing is an important constraint for assessing the impact of the
Grand Tack, dynamical shake-up, and other sculpting mechanisms.
Calculations of planet formation for the inner solar system predict
that Mars-size protoplanets form on million-year time scales at 1.5~AU
\citep{chambers1998, kb2006, raymond2009}. Radiometric data support
this idea, indicating a fully assembled Mars within $\sim$4~Myr
\citep{dauphas2011b}.  With depletion of the reservoir of gas in
$\sim$1--5~Myr \citep{haisch2001, kk2009, wang2017}, the gas giants
probably form before Mars reaches its final mass.  Thus, within a few
million years of the collapse of the solar nebula, the gas giants were
largely in place, the gas disk was diminished, and Mars was near
completion.

Compared to these evolutionary time scales, the Grand Tack model
operates on a tight schedule \citep{walsh2011}. In this picture, the
gas giants fully form within a million years and begin to migrate
within an undepleted gaseous disk.  After Jupiter and Saturn begin
their return trip through the gaseous disk --- the ``tack'' --- the
gas beyond the asteroid belt must dissipate rapidly to end outward
migration.  The outcome of this process is a reduced surface density
of solids beyond 1~AU, leaving a small Mars and a dynamically excited,
depleted asteroid belt.  While compelling, this picture requires
careful synchronization between gas giant formation, the evolution of
the gas disk, and the growth of solids at 1--3~AU
\citep{raymond2014b}.

In contrast, the main timing constraints for dynamical shake-up are
that (i) the gas disk vanishes inside 5--10~AU after the formation of
Jupiter and (ii) the disk inside the Kuiper belt dissipates on time
scales of millions of years \citep{naga2000}. The dissipation rate
governs (i) how quickly the $\nu_5$ and other resonances sweep through
the inner solar system and (ii) how long individual objects experience
the resonance.  A resonance that sweeps rapidly may not shake up
anything; a slowly moving resonance can lead to dynamical
ejections. For disk dissipation time scales in the range favored by
observations \citep[e.g.,][]{haisch2001, kk2009, bell2013,
  pfalzner2014, ribas2015, wang2017}, resonances sweep through slowly
enough to make an impact with little sensitivity to the way the disk
dissipates \citep{naga2000}.  Therefore, a dynamical shake-up almost
certainly played an important role in the formation of the solar
system.

Here, we consider the possibility that the low mass of Mars and the
asteroid belt might result directly from the dynamical shake-up,
without the need for large-scale migration of the gas giants. Although
previous studies \citep{naga2005, obrien2007, nagasawa2007,
  thommes2008} showed promising results, both the shake-up and the
formation of Mars were thought to have occurred after $\sim
5$--10~Myr.  Armed with more recent estimates of Mars' formation time,
we propose that the shake-up happened earlier, when planet formation
in the inner solar system was far from complete.  To assess the impact
of this early shake-up requires tracking the evolution of
\textit{planetesimals}, including the physics of coagulation.
Collisional damping is also critical to the outcome, as it counteracts
dynamical excitations from sweeping resonances.  Our new contribution
is a set of calculations that include these effects, performed with
our planet formation code, \orchestra\ \citep{bk2006, bk2011a,
  bk2013}.

We organize this paper as follows. In \S\ref{sect:orbdyn} we consider
orbital dynamics in the inner solar system, including the nature of
particle orbits in the presence of outer gas giants and the mechanics
of a sweeping resonance driven by an evolving gas disk.  We
incorporate these phenomena in our planet formation code, summarized
in \S\ref{sect:coag}.  In \S\ref{sect:num}, we present the results
from two extensive multiannulus simulations that compare rocky planet
formation with and without a dynamical shake-up.  In our conclusion,
\S\ref{sect:done}, we summarize the advantages and limitations of
dynamical shake-up for explaining the low mass of Mars. We also
discuss the implications of our results for terrestrial planet
formation elsewhere in the Universe.

\section{Orbital dynamics in the inner solar system}
\label{sect:orbdyn}

We first consider the orbital dynamics of growing planetesimals and
protoplanets within the context of the shake-up model. Our motivation
here is to understand how to include the gravitational effects of
giant planets and a gas disk in our planet formation code. First, we
focus on Jupiter, which produces a significant time varying
gravitational potential that affects orbital motion throughout the
solar system. Outside of its Hill sphere, however, there are sets of
``most-circular'' orbital paths that enable disks of particles to be
dynamically cold. Along these orbits, stirring by Jupiter is
negligible. Thus, planet formation in the inner solar system can
proceed as if Jupiter were not there. A similar situation exists for
binary stars \citep{bk2015tatooine}.

The presence of a massive gaseous disk in the Sun--Jupiter binary
complicates this simple picture. From secular perturbation theory,
orbital precession driven by Jupiter and the disk can lead to secular
resonances that excite eccentricities of planetesimals and
protoplanets. The resonance location changes along with the disk
potential as the disk evolves.  Although aerodynamic drag and
dynamical friction with the gas can (i) damp eccentricity and (ii)
remove small particles from the system, strong secular resonances that
sweep through the disk can generate large orbital eccentricity $e$
that may persist even after the disk vanishes.

In this section, we cover orbital dynamics in the presence of an outer
giant planet, an evolving gas disk, and the sweeping resonances they
produce. Following \citet{naga2005} and \citet{thommes2008}, we focus
here on the strong $\nu_5$ apsidal resonance, although nodal
resonances may also contribute to a shake-up
\citep[e.g.,][]{haghighipour2016}.  We include several examples that
illustrate how a sweeping resonance might have affected protoplanetary
orbits in the terrestrial zone of the young Sun.

\subsection{Orbit solutions and ``most circular'' paths}

To quantify the orbits of small bodies in the inner solar system,
we first consider the isolated influence of Jupiter, treating it as a
binary partner to the Sun.  Following the theory of \citet{lee2006}
and \citet{leung2013}, we assume that particle orbits make small
excursions about a guiding center on a circular track around the Sun.
We then solve equations of motion to linear order in these excursion
distances.  \citet{lee2006} and \citet{leung2013} originally focused
on circumbinary systems where the eccentricities of all objects,
including the binary, are low. Here we extend this theory to the
\circpriintro\ case.

Using the prescription in Appendix~\ref{appx:mostcirc}, we obtain
orbit solutions for tests particles interior to
Jupiter's orbit.  For orbital excursions that are small compared to
Jupiter's semimajor axis $\ajup$ and for low orbital eccentricities
$e$, orbit solutions are
\begin{eqnarray}\label{eq:orbsolnr}
R(t) & \approx & \rgc \left[1 - \efree \cos(\kappae t+\psi_e) 
 - \eforce \cos(\ngc t) - F(t,\rgc,)\right],
\\
\phi(t) & \approx & \ngc \left[1 + 
\frac{2\efree}{\kappae} \sin(\kappae t+\psi_e) 
+ \frac{2\eforce}{\ngc} \sin(\ngc t) + G(t,\rgc)\right],
\\
z(t) & \approx & \rgc i \sin(\kappai t +\psi_i)
\end{eqnarray}
where $R$, $\phi$ and $z$ are cylindrical coordinates with the origin
at the Sun and with Jupiter in the $z=0$ plane. We define the other
variables and functions according to how they contribute to the test
particle's motion:
\begin{enumerate}
\item  The radius $\rgc$ defines the circular path of the particle's
 guiding center, which has an orbital frequency $\ngc$;

\item The ``free eccentricity'' $\efree$ and inclination $i$ describe
  motion in the epicyclic approximation of Keplerian orbits around the
  Sun, with the angular frequencies $\kappae$ and $\kappai$,
  respectively, which differ slightly from $\ngc$;

\item The ``forced eccentricity'' $\eforce$ corresponds to epicyclic
  motion in the Sun-Jupiter plane, driven at angular frequency $\ngc$
  by Jupiter; and

\item The functions $F$ and $G$ quantify additional modes of
  oscillation that depend on harmonics of Jupiter's mean motion and
  the synodic frequency of Jupiter and the test particle.
\end{enumerate}
The remaining variables, $\psi_e$ and $\psi_i$, are phase angles
defined to apsidally align the test particle with Jupiter at $t = 0$.


Because a test particle's epicyclic frequency $\kappae$ and vertical frequency
$\kappai$ differ from the mean motion $\ngc$, its
argument of perihelion and ascending node precess
(Appendix~\ref{appx:mostcirc}).  For $\ajup$ = 5.2~AU, the apsidal
precession rate is
\begin{equation}
\label{eq:omprejup}
\ompresJ \equiv \ngc - \kappae \approx 
\frac{3}{4}
\nkep
\frac{\Mjup}{\Msun}\frac{a^3}{\ajup^3}
\approx 
36\times \aAU^{3/2} \ \textrm{rad/Myr,}
\end{equation}
where we have replaced $\rgc$ with orbital distance $a$,
and the subscript (${p,J}$) specifies precession of the test particle ---
representing a 
planetesimal or protoplanet --- as a result of Jupiter's influence.
The corresponding apsidal precession time is
\begin{equation}\label{eq:PpresJ}
T_{p,J} \approx 0.17 \aAU^{-3/2} \ \textrm{Myr.}
\end{equation}
This time scale is (i) short compared to the 3--5~Myr lifetime of the
gaseous disk \citep[e.g.,][]{bell2013,ribas2015} and (ii) comparable
with the time scale for Jupiter and Saturn to migrate through the disk
\citep[e.g.,][]{walsh2011}.

The forced eccentricity imposed by Jupiter on a small body orbiting
interior to the gas giant is
\begin{equation}\label{eq:eforce}
\eforce
\approx 
\frac{5}{4}\frac{a}{\ajup} \ejup
\approx 0.012 \aAU \ejupNow 
\ \ \ \ \ \textrm{($a\ll \ajup$)},
\end{equation}
where $\ejup$ is Jupiter's eccentricity.  Unlike the free eccentric
motion, the forced eccentric mode remains apsidally aligned with
Jupiter's orbit.

\begin{figure}[htb]
\centerline{\includegraphics[width=7.0in]{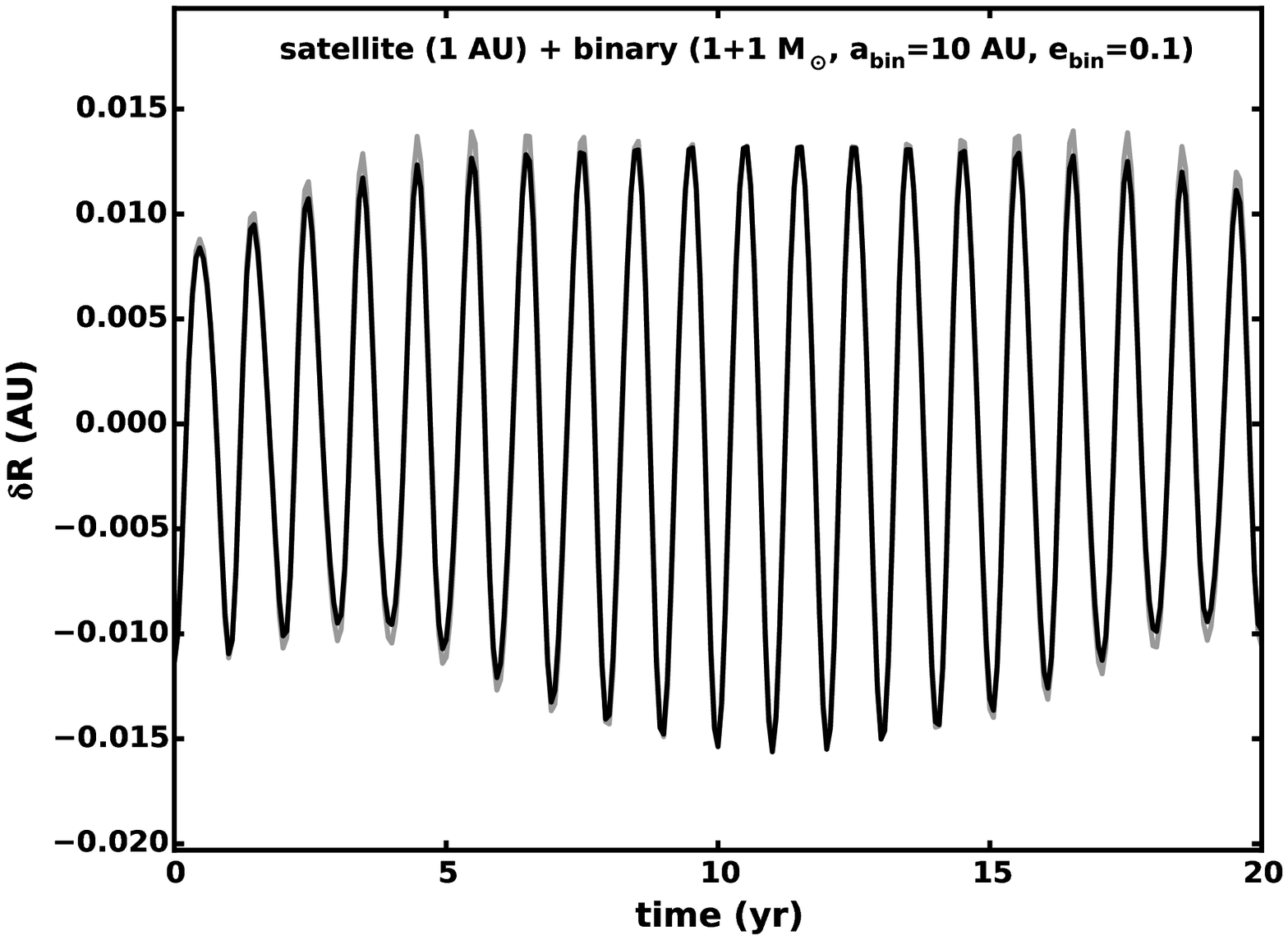}}
\caption{\label{fig:mostcircstarz} Radial position of a test particle
  (a ``satellite'', such as a planetesimal or protoplanet) on a most
  circular orbit about a star in a stellar binary system.  The two
  stars have mass equal to 1~\Msun\ and orbit with semimajor axis $a$
  = 10~AU and eccentricity $\ebin = 0.1$.  The particle's average
  distance from its primary host is $\rgc=1$~AU.  The radial
  fluctuations show the rapid forced eccentric oscillations plus
  slower modulations from the binary's epicyclic motion.  For the
  Sun-Jupiter system, the rapid radial oscillations are similar to the
  curve in the plot, but the slower modulations are not evident.  }
\end{figure}

From the orbit solutions, planetesimal and protoplanets follow ``most
circular'' paths when $\efree$ goes to zero. These trajectories are
generalizations of circular orbits around an isolated star that allow
for ebb and flow in response to gravitational perturbations from
Jupiter \citep[e.g.,][]{lee2006, youdin2012, bk2015tatooine}.
Figure~\ref{fig:mostcircstarz} illustrates most-circular orbit
solutions for an \circpri\ planetesimal in the extreme case of an
equal-mass stellar binary.  To achieve these orbits, swarms of
planetesimals collisionally damp, ridding themselves of free eccentricity.
The time scale for this dynamical cooling process is limited by the
precession time of the free eccentricity, Equation~(\ref{eq:PpresJ}),
as well as the collision time.  Furthermore for settling to occur, the
binary eccentricity must be modestly low ($\ejup \lesssim 0.5$), so
that most-circular orbits are nested and non-intersecting (see
below).\footnote{An odd feature of the most-circular path formalism
  for an \circpri\ system is that the magnitude of the forced
  eccentricity is independent of the secondary mass. Thus, if
  Jupiter's mass were reduced to some arbitrarily small value, a
  hypothetical disk of dust would settle into exactly the same
  most-circular configuration as before. However, the time it would
  take for a disk to transition from a purely circular configuration
  to a most-circular one is roughly limited by the apsidal precession
  rate \citep[Equation~(\ref{eq:omprejup});
    see][]{marzari2000,thebault2002,thebault2006,silsbee2015a}.  If
  Jupiter were less massive than the Moon, then the settling time at
  1~AU would exceed the age of the solar system.}

\subsection{Orbits in a massive gas disk}

To include a gaseous disk in our planet formation calculation, we
adopt a simple axisymmetric model, with surface density
\begin{equation}\label{eq:Sigma}
  \Sigma(a, t) = \Sigma_0 \exp(-t/\tau) \left(\frac{a}{a_0}\right)^{-n} 
\ \ \ \ \ \ \ \ \ \text{($\ain < a < \aout$ [0 otherwise])},
\end{equation}
where $a$ is the distance from the Sun, $t$ is time, $\Sigma_0$, $n$
and $\tau$ are constants, $\ain$ and $\aout$ are inner and
outer disk radii, and $a_0 \equiv 1$~AU.  
By setting $\Sigma_0 = 2000$~g/cm$^2$ and $n = 1$, we adopt a model
similar to the Minimum Mass Solar Nebula \citep[MMSN;][]{weiden1977b,
  hayashi1981}, which is comparable in mass with some disks observed
around young stars \citep[e.g.,][]{and2009, and2010, dent2013,
  and2013}.  Choosing a time scale $\tau \sim 1$--3~Myr reproduces
observed disk dissipation rates if the disk erodes ``uniformly'' over
its surface \citep{haisch2001, will2011}.  Alternatively, the disk may
erode from the inside out with an inner disk radius
\begin{equation}
  \ain(t) = \ain(0) + \openrate t,
\end{equation}
where $\openrate$ is a constant expansion rate. Theory
\citep{owen2012, clarke2013} and observations of transition disks
\citep{calvet2005, currie2008, and2011} suggest that $\openrate
\gtrsim 10$~AU/Myr \citep[see also][]{owen2016}. Here, our disk models
either decay uniformly with a finite $\tau$ or they have an expanding
inner edge, but not both. In all cases, we fix $\aout$ to be 100~AU.
Results described below are insensitive to this choice.

For the vertical structure of the disk, we set the characteristic
density in the disk as $\rho \sim \Sigma/h$ with vertical scale height
\begin{equation}
h = h_0 \left(\frac{a}{a_0}\right)^{5/4},
\end{equation}
and $h_0 = 0.04$~AU \citep{kh1987, and2007, and2009}.

To model the spatial distribution of solids, we assume that the
surface density $\Sigma_{s,0} = 10$~g/cm$^2$ and $n = 1.5$ for
material inside the snow line. Solids in the terrestrial zone then
comprise about 0.5\% of the initial disk mass.  Solid particles have a
size-dependent scale height, which depends on the relative importance
of settling, gas drag, and gravitational stirring
\citep{weiden1980,gold2004,chiang2010}.  For simplicity, we assume
that the surface density of solids does not deplete with the gas, but
evolves independently of the gas disk.

Finally, we assume that Jupiter and Saturn create and maintain gaps in
both the gas and solid particle disks. Thus we set the surface density
to zero in annuli of width 1~AU, centered on each of the gas giants.
This point is important because the sense of apsidal precession of the
gas giants themselves depends on whether they are embedded in the
disk. We discuss this issue next.

\subsubsection{Disk gravity}\label{sssect:diskgrav}

The non-Keplerian potential of a massive disk also results in orbital
precession.  Aside from contributing to the location of secular
resonances, the time-varying gravitational potential of the disk as it
dissipates is the key to resonant sweeping and the dynamical shake-up
model.  While we use a grid-based method to handle disk gravity in our
orbit solvers \citep{bk2011a}, analytical estimates offer some insight
into the impact of a disk on solid objects.

For a simple, power-law disk with no gaps, the potential at distance 
$a$ within the bulk of the disk is
\begin{equation}\label{eq:diskpotential}
  \Phi_d(a) \approx
\begin{cases}
2 \pi G\Sigma_{0} \exp(-t/\tau) a_0 \log(a/a_0) 
& \ \ \ \ \ \ \ \ \textrm{($n=1$; gas)}\\
-8.8\pi G\Sigma_{0} \exp(-t/\tau) a_0 (a/a_0)^{1.5} 
& \ \ \ \ \ \ \ \ \textrm{($n=1.5$; solids)}
\end{cases}
\end{equation}
\citep[Appendix A of][]{bk2011a}.  This potential induces apsidal
precession at a rate of
\begin{equation}\label{eq:ompredisk}
\ompresd \approx -\pi \frac{a^2 \Sigma(a,t)}{M} \nkep
\ \ \ \ \ \ \ \ \ \ \ 
\textrm{(extended power-law disk),}
\end{equation} 
where the subscript (${p,d}$) signifies the precession of a planetesimal
(or protoplanet) caused by the disk (Appendix~\ref{appx:secular},
Equation~(\ref{eq:omprediskgen})). This expression is exact for $n=1$
and is accurate to about 10\% for $n = 1.5$ \citep[e.g.,][]{raf2013,
  silsbee2015a}.  The negative sign indicates that a planetesimal
embedded in a disk has an apside that precesses in a sense opposite to
its orbital motion.  The precession period is
\begin{equation}\label{eq:Pprediskuni}
T_{p,d} \approx 0.14 \aAU^{5/2} \SatXXcgs^{-1} \ \textrm{Myr}
\ \ \ \ \ \ \ \ \ \ \ \textrm{($\ain\ll a\ll\aout$, $n = 1$)}.
\end{equation}
We choose numerical factors to highlight conditions where the
precession from the disk is comparable to the precession induced by
Jupiter (Equation~(\ref{eq:PpresJ})). Similar precession periods lead
to resonances and the dynamical shake-up.

The disk gravity affects the gas giants as well, but because these
planets can clear gaps, the nature of their precession is
different. For Jupiter centered in a 1~AU full-width gap, the
precession rate is
\begin{equation}
\ompreJd = 1.25\pi \frac{\ajup^2 \Sigma(\ajup)}{M}\omjup
\end{equation}
where $\omjup$ is the angular speed of Jupiter's orbit.  We obtain the
numerical coefficient from numerical experiments (see below) using a
disk with a gap and a finite scale height. This value is within a
factor of about two of the analytical estimate of \citet[Equation~(23)
  therein]{naga2005} for an infinitesimally thin disk.

To describe scenarios in which the circumsolar gas erodes from the
inside out, we estimate the precession rate of a planetesimal or
protoplanet using Equation~(\ref{eq:omprejup}) for the orbit-averaged
precession induced by a planetary perturber. By substituting $2\pi
\Sigma a^\prime da^\prime$ for the perturber mass and integrating over
orbital distance $a^\prime$,
\begin{equation}
\ompresd \approx \frac{3}{4} \nkep
\frac{\pi \ain^2\Sigma(\ain)}{\Msun} \frac{a^3}{\ain^3}
\ \ \ \ \ \ \ \ \ \ \
\textrm{($a\ll \ain$, $n=1$; {inner cavity})}.
\end{equation}
This expression is valid when the inner edge of the disk is far beyond
the orbit of the planetesimal or protoplanet.  When
resonances sweep through the terrestrial zone, the inner edge of the
disk is well outside of Saturn's orbit. This expression is then a useful
approximation. The precession period is
\begin{equation}
T_{p,d} = 0.19 \aAU^{-3/2} \SinCCcgs^{-1} \ainXAU \ \textrm{Myr.}
\ \ \ \ \ \ \ \ \ \ \ \textrm{($a\ll\ain$; $n = 1$)},
\end{equation}
where the choice of numerical factors reflects the fact that the
surface density of a MMSN disk at 10~AU is 200~g/cm$^2$ when $n$ = 1.

In comparing an extended power law disk and a disk with an inner
cavity, a planetesimal or protoplanet within a disk cavity or gap
precesses in the same sense as its orbital motion, while an object
embedded within a power-law disk precesses in the opposite
sense. Despite these differences, in our models the magnitude of
$\ompresd$ generally decreases monotonically with $a$.  Furthermore,
the effect of the disk on the precession of a planetesimal in the
terrestrial zone relative to Jupiter is fairly insensitive to the way
the disk dissipates in our models.

For disks with finite spatial extent and gaps, we use a numerical
approximation to quantify the gravitational acceleration.  After
dividing the disk into 2500 annuli spanning orbital distances from
0.1~AU to 100~AU, we evaluate the disk gravity at each annulus and
derive real-time acceleration evaluations by interpolation
\citep[see][]{bk2011a}. In this way we can accommodate a wide range of
disk surface density profiles. Our code can perform these calculations
in 3-D around an axisymmetric disk.  However, our focus here is on
eccentricity pumping, and we work only in the disk midplane, defined
to have no inclination with respect to the Sun-Jupiter binary.

\subsubsection{Resonant excitations}
\label{subsubsect:rez}

Even with only modest eccentricity, Jupiter continuously pumps the
eccentricity of objects throughout the solar system. The pumping
rate is
\begin{equation} \label{eq:epump}
\epumprate \approx \frac{15}{16}
\nkep\frac{\Mjup}{\Msun}\frac{a^4}{\ajup^4}\,\ejup
\approx 0.44\times \aAU^{5/2} \ejupNow  \ \textrm{Myr$^{-1}$}
\end{equation}
(Appendix~\ref{appx:secular}, Equation~(\ref{eq:B})).  If a
planetesimal experiences apsidal precession relative to Jupiter, then
its orbit can be stable, despite the pumping: Precession constantly
reorients the planetesimal's orbit, preventing the eccentricity from
coherently building up. The forced eccentric motion arises from a
balance between pumping and precession; the more rapid the relative
precession, the smaller the forced eccentricity.

By modulating the precession rates of both Jupiter and a planetesimal, a
massive gas disk complicates this picture. The forced eccentricity
becomes (Appendix~\ref{appx:secular})
\begin{equation}\label{eq:eforcepJd}
\eforce \approx \frac{5}{4}
\left[
\frac{\ompresJ}{(\ompresJ + \ompresd) - \ompresJ}
\right]
\frac{a}{\ajup}\ejup, 
\end{equation}
where $\ompreJd$ is the apsidal precession rate of Jupiter from the
disk. Thus, the denominator describes the precession rate of the
planetesimal relative to Jupiter.

This form of the forced eccentricity has implications for the ability
of a swarm of planetesimals to settle on most circular orbits.  Near a
secular resonance, where planetesimals and Jupiter precess at the same
rate, the forced eccentricity diverges and orbits necessarily cross.
Thus, there are regions near the resonance where solid material cannot
settle onto most circular orbits. To prevent orbit crossing,
\begin{equation}\label{eq:nocross}
 \left|\eforce + a \frac{d\eforce}{da}\right| < 1 ~ ,
\end{equation}
which comes from the requirement that the perihelion distance of
planetesimals on adjacent most-circular orbits do not overlap.  In
regions where this condition is met, away from secular resonances,
non-intersecting most circular orbits exist, and small solids can
settle to a dynamically cold state.

\subsubsection{Resonant sweeping}

We derive more general circumstances for secular resonances by
expressing the precession rate of a planetesimal (or protoplanet)
relative to Jupiter as
\begin{eqnarray}\label{eq:relpresJ}
\ompres - \ompreJ 
& = & 
\ompresJ + \ompresS + \ompresd - \ompreJS - \ompreJd,
\end{eqnarray}
where $\ompresS$ and $\ompreJS$ are the precession rates that Saturn
imposes on the planetesimal and Jupiter, respectively.  When $\ompres
= \ompreJ$, the planetesimal experiences the $\nu_5$ secular resonance
with Jupiter.  The location of the resonance varies with the evolution
of the disk.  In a very massive disk, planetesimal orbits precess so
rapidly that no resonance exists within the orbit of Jupiter
($|\ompres| \gg |\ompreJ|$ for all $a < \ajup$).  As the disk
dissipates, the planetesimal precession slows, until the effect of
precession from Jupiter's gravity kicks in. Because Jupiter causes
precession most strongly at orbital distances closest to it, the
$\nu_5$ resonance appears first beyond the terrestrial zone, and
sweeps inward as the disk fades away, settling just beyond the orbit
of Venus.

\begin{figure}[htb]
\centerline{\includegraphics[width=7.0in]{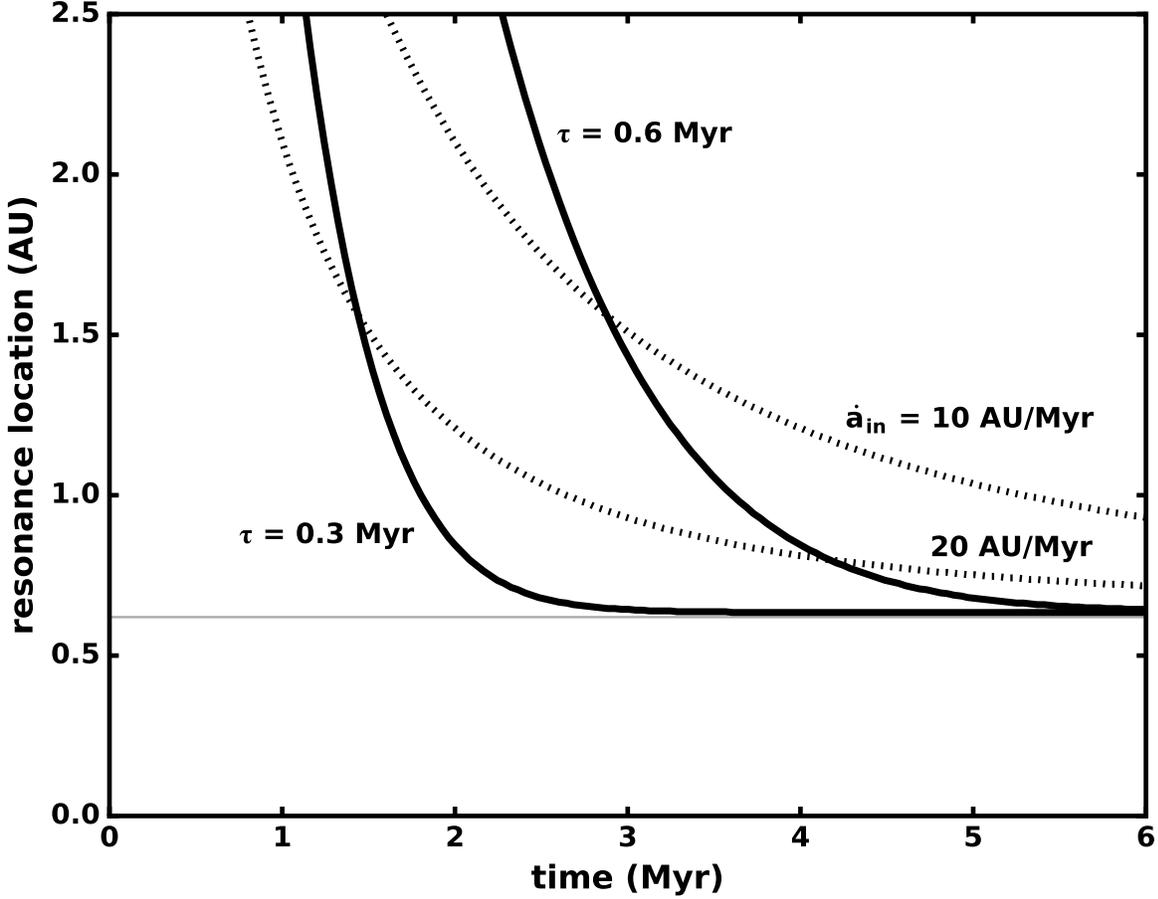}}
\caption{\label{fig:rezsweep} The location of the $\nu_5$ secular
  resonance in two types of disk evolution models with Jupiter (Saturn)
  in orbit at $a_J$ = 5.20 AU ($a_S$ = 9.58 AU).  The solid curves show 
  the orbital distance where (i) a test particle's apsidal precession
  rate matches Jupiter's and (ii) the gas disk depletes uniformly with 
  depletion time scale $\tau$. The dotted curves correspond to inside-out 
  depletion of the disk, where $\openrate$ is the expansion rate of the 
  disk's inner edge. The curves come from analytical approximations that 
  include the gravity of the disk, Saturn's influence, and a Jupiter which
  clears a gap in the disk if the disk's inner edge is within Jupiter's orbit.  
}\end{figure}

Figure~\ref{fig:rezsweep} illustrates the evolving location of the
$\nu_5$ resonance in the uniform depletion and inside-out disk
dissipation models. To estimate this location, we calculate
eigenfrequencies of precession for the Jupiter-Saturn system in
secular theory \citep[e.g.,][]{heppenheimer:1980, naga2000}, adopting
the present-day orbital configuration of the two gas giants ($a_J$ =
5.20~AU; $a_S$ = 9.58~AU).  The figure shows that the timing and
location of the sweeping resonance depends on the form of disk
dissipation. Slower dissipation yields a more slowly sweeping
resonance.  The resonance sweeps through the inner solar system more
slowly with inside-out dissipation than with homologous disk
depletion.

\begin{figure}[htb]
\centerline{\includegraphics[width=7.0in]{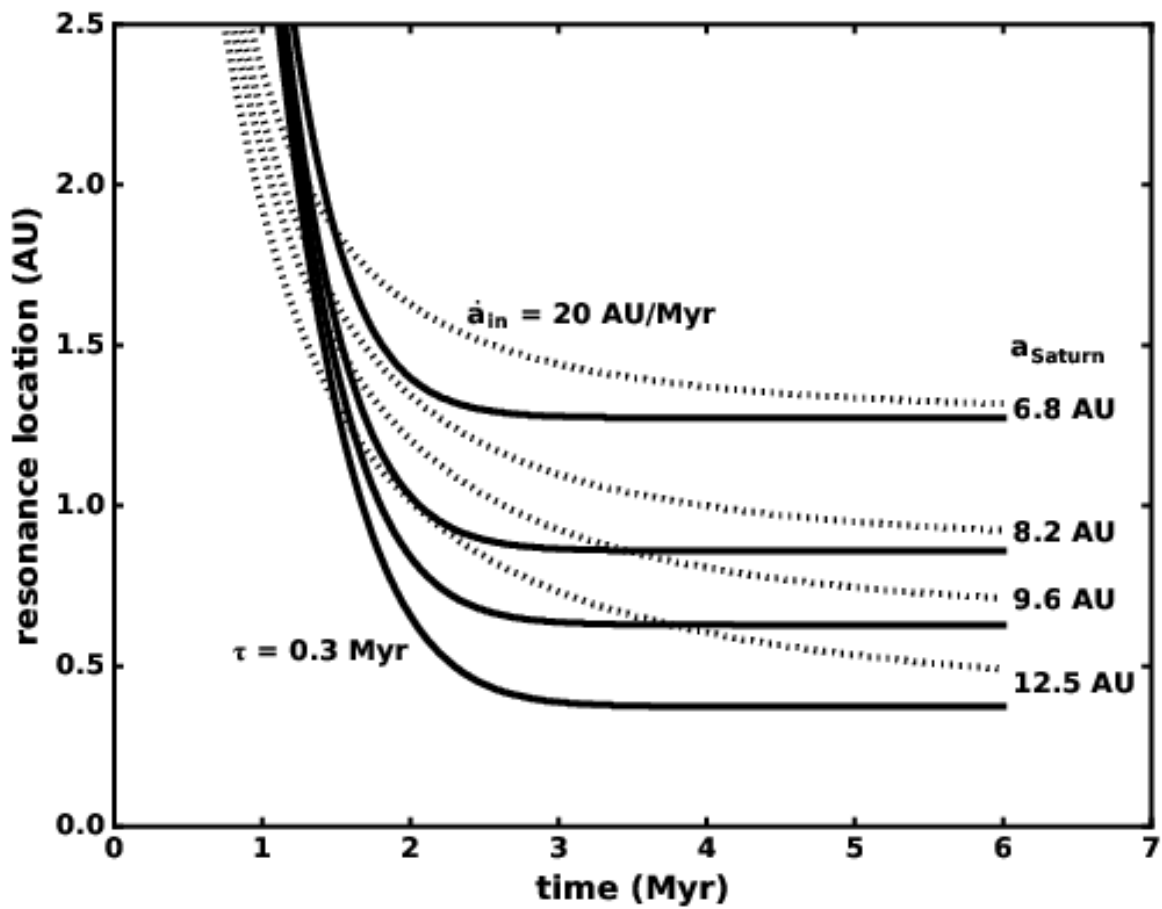}}
\caption{\label{fig:rezsweepnewsaturn} The location of the $\nu_5$
  secular resonance for several different orbits of Saturn.  As in the
  preceding Figure, the solid curves correspond to uniform disk
  depletion, while the dotted curves correspond to inside-out
  depletion of the disk. Parameters for each model are shown. For each
  disk depletion model, the position of the resonance is calculated
  with Saturn at one of four positions: Two positions correspond to
  the 3:2 and 2:1 Jupiter-Saturn commensurabilities near 6.8~AU and
  8.2~AU respectively, while the third is near Saturn's present-day
  orbit. The fourth position for Saturn is chosen as an example when
  the Jupiter-Saturn orbital separation was greater than the present
  day value. In all cases, Jupiter is at an orbital distance of
  5.2~AU.  }\end{figure}

Figure~\ref{fig:rezsweepnewsaturn} highlights the role of Saturn in
setting the pace and timing of resonance sweeping.  Although the final
position of the resonance is sensitive to Saturn's orbital distance,
the timing of sweeping beyond 1~AU is rather insensitive to $a_S$.  In
inside-out dissipation with Jupiter and Saturn near the 3:2 orbital
commensurability, the $\nu_5$ resonance settles at roughly 1.3~AU; the
sweep rate is then comparatively slow as the resonance crosses Mars'
orbit. The disk model then mimics a more slowly evolving disk. In
calculations with larger $a_S$, the resonance sweeps rapidly through
Mars' orbit independent of the exact value of $a_S$.

These results demonstrate that outcomes of dynamical shake-up
calculations are not sensitive to $a_S$. Whether Saturn orbits in its
current location as in classical models
\citep[e.g.,][]{haghighipour2016} or in a more compact configuration
with Jupiter as in the Nice model \citep{tsig2005}, the $\nu_5$
resonance passes through the asteroid belt and the orbit of Mars.  The
pace and timing of resonance sweeping are more sensitive to the form
and timing of disk evolution than the orbit of Saturn.

\subsubsection{Dynamical friction and eccentricity damping}

In addition to generating an overall axisymmetric potential, the gas
disk produces gravitationally important density wakes as it responds
to the local gravity of large planetesimals, protoplanets and
planets. By creating these wakes, a massive object experiences
eccentricity damping and radial migration through the disk
\citep{lin1979a, gold1980, lin1986, arty1993, ward1997}. Eccentricity
damping arises because of the ``downstream'' density enhancements
caused by Rutherford scattering as a planet plows through the gas
disk. Once it has circularized, the planet feels torque from the gas
in Keplerian flows streaming by, flowing in one direction on interior
orbits and the other direction on exterior orbits. The slight torque
differential causes the planet to drift radially. The exact mechanism
depends on whether the planet is massive enough to clear a gap in the
disk \citep{ward1997}. Small planets unable to generate a gap
experience type I migration, which typically is slow in the
terrestrial zone compared to the disk lifetimes considered here. More
massive planets that open a gap undergo Type II migration, which
operates on viscous time scales that are short compared to the disk
depletion time scale in a massive disk.


We focus first on eccentricity damping.  The damping time scale is
\begin{eqnarray}\label{eq:edamp}
\tauedamp & \equiv & 
\frac{e}{\dot{e}}  = \frac{1}{\nkep}\ \frac{\Msun}{\mp}
\ \frac{\Msun}{a^2\Sigma}\ \frac{h^4}{a^4}
\nonumber
\\
\ & \approx &
0.6 \aAU^{1/2}\mOpIMe^{-1}\SatXXcgs^{-1}\hOpOIV^{5} 
\ \textrm{Myr}
\end{eqnarray}
\citep{arty1993, agnor:2002}. When the surface density of the disk
is close to its initial value, $\sim$ 1000--2000~g~cm$^{-2}$, $e$
damps quickly for planets with $m \gtrsim$~0.01~\mearth. When the
$\nu_5$ resonance sweeps through the inner solar system, 
$\Sigma \sim$ 20~g~cm$^{-2}$. Over the remaining lifetime of the 
disk, damping is important only for planets with masses comparable to 
or larger than Mars.

Gravitational wakes also produce radial drift \citep{lin1979a,
  gold1980}.  For objects in the terrestrial zone with insufficient
mass to create a gap in the gas disk, the slow drift time scale for
Type~1 migration is
\begin{eqnarray}
\nonumber
\tauadrift
& \equiv &
\frac{a}{\dot{a}}  \approx \frac{1}{\nkep}\ \frac{\Msun}{\mp}\ 
\frac{\Msun}{a^2 \Sigma}\ \frac{h^2}{a^2}
\ \ \ \ \ \ \ \ \ \ \ \ \ \ \ \ \textrm{(Type I)}
\\
\ & \approx & 400 
\mOpIMe^{-1}\SatXXcgs^{-1}\hOpOIV^2 \ \textrm{Myr}
\end{eqnarray}
\citep{tanaka2002}. In an evolved gas disk, Type~I migration is
not important for planet formation in the inner solar system.

Planets with masses large enough to carve a gap in the gas disk can
experience Type II migration \citep{lin1986, ward1997}.  For these
objects, the radial drift time is 
\begin{eqnarray}
\nonumber
\tauadrift
&  \approx & 0.1 \frac{a^2}{\alpha\nkep h^2} \frac{\mp}{\pi a^2\Sigma} 
\ \ \ \ \ \ \ \ \ \ \ \ \ \ \ \ \ \textrm{(Type II)}
\\
\label{eq:typeII}
\ & \approx & 7.3 \alphaOpOOOI^{-1} 
\hOpOIV^{-2} \mIMjup \SOhMMcgs^{-1} \exp[0.25 t/\tau] \ \textrm{Myr}
\end{eqnarray}
where $\alpha$ is the disk viscosity parameter
\citep[see][]{dangelo2003, papa2007, alex2009, duffell2014,
  durmann2015, tanigawa2016}. Here, we choose to evaluate the disk
surface density at four e-folding times, just prior to the resonant
sweeping in the asteroid belt \citep[cf.][]{naga2005}.  This
expression applies only to planets that are Saturn-mass or larger;
smaller planets have Hill radii smaller than the disk scale height and
cannot clear a gap \citep[e.g.,][]{crida2006}.  For Jupiter in a
massive disk, drift times are short, as in the Grand Tack model.  When
the $\nu_5$ resonance sweeps through the inner solar system, however,
the low surface density limits the radial drift of Jupiter through the
disk. With a drift time longer than the disk lifetime, Jupiter must be
close to its present location during the dynamical shake-up.

At the onset of dynamical shake-up, the migration time scale for
Saturn is shorter than for Jupiter. However, a drift time of 2~Myr is
still longer than the short disk dissipation time scales considered
here. Even if Saturn drifts during the shake-up, the pace and timing
of resonance sweep through the asteroid belt and across the orbit of
Mars are not expected to change (see
Fig.~\ref{fig:rezsweepnewsaturn}). While faster, Type~III migration is
possible for Saturn when the disk is massive \citep{masset2003}, rapid
drift is unlikely when the gas disk surface density is as low as in
the scenarios considered here \citep[e.g.,][Fig.~6,
  therein]{papa2007}. While migration may be essential for
establishing the orbital configuration of Saturn and Jupiter at early
times \citep[e.g.,][]{masset2001, morbycrida2007, zhang2010}, it is
not an important consideration during dynamical shake-up.


Although a lower mass planet may not experience steady inspiral through 
Type I or Type II migration during dynamical shake-up, it can drift 
inward if it initially has some eccentricity.  As its eccentricity damps, 
the planet loses energy and drifts radially inward \citep{ada76, 
thommes:2003, naga2005}. However the drift time is roughly the eccentricity 
damping time scaled by $e^{-2}$. As long as damping keeps the eccentricity 
low, radial drift is insignificant.  The scenarios considered here have 
short episodes of eccentricity pumping; radial drift is then small. We 
confirm this behavior in \nbody\ simulations and therefore omit it from 
consideration. In contrast, \citet{naga2005} and \citet{thommes2008} consider
models with episodes of eccentricity pumping and damping that last for
millions of years. Radial drift is an important feature in those scenarios.

\subsubsection{Gas drag}

Aerodynamic drag also impacts the orbits of solid particles in the
disk \citep[e.g.,][]{chiang2010}.  The behavior of solids depends on
the mean free path of gas molecules \citep{ada76, weiden1977a}:
\begin{equation}
\mfp \approx \frac{1}{\sigma \ngas}
\approx 100 \aAU^{5/4} \SatXXcgs^{-1} \hOpOIV \ \textrm{cm},
\end{equation}
where $\sigma$ is the cross section of gas molecules, $\ngas$ is their
number density, and the numerical factors in the rightmost expression
are suggestive of circumstellar nebula conditions after the gas giants
have formed.  If particles are comparably sized or smaller than
$\mfp$, then molecular collisions dominate the dynamics according to
Epstein's Law.  Larger particle sizes or high relative speeds can
cause different behavior: Slowly moving particles larger than $\mfp$
experience a viscous force that is proportional to the relative speed
--- Stoke's Law.  Particles with a radius $\rp \gg \mfp$ or which have
a speed that greatly exceeds the sound speed of the gas produce
turbulence and feel a force proportional to the square of the relative
speed.

To quantify the importance of gas drag, we consider the stopping time
of a particle, defined as
\begin{equation}
\tstop \equiv \left|\dv/\aaero\right|,
\end{equation}
where $\aaero$ is the acceleration from aerodynamic drag and $\dv$ is
the relative speed between the particle and the gas.  An
astrophysically important value for $\dv$ is the difference between
the Keplerian orbital speed of a particle, $\vkep$, and the circular
speed of the gas disk, $\vgas$, which is slower because of pressure
support.  From simple assumptions about the gas pressure
\citep[e.g.,][]{yk2013},
\begin{equation}
\vkep - \vgas 
\equiv
\eta \vkep,  
\ \ \ \ \textrm{with} \ \ \ \
\eta \sim 
(h/a)^2 \sim 10^{-3}, 
\end{equation}
the ``headwind'' felt by the particle has a 
Mach number of 
\begin{equation}
\ma = \eta\frac{\vkep}{\cs} \sim \frac{h}{a} = 0.04 {\hOpOIV}\aAU^{1/4},
\end{equation} 
where we have used a sound speed $\cs \sim h\vkep/a$.  With this Mach
number and the mean free path in the disk, we can determine the
stopping time as a function of particle size.

When the gas density drops to about 1\% of its MMSN value, the
stopping time becomes comparable to a dynamical time for
centimeter-size particles.  Just as for meter-size particles at
earlier epochs \citep[the `meter-size barrier';][]{weiden1977a,
  johansen2007}, these solids cannot become fully entrained in the gas
and are doomed to spiral in toward the Sun on a time scale of $\sim
1/\eta$ orbital periods.  Under these conditions, smaller dust
particles are entrained in the gas.  Larger planetesimals are less
affected by aerodynamic drag. For example, objects with a size of
roughly 10~km are in the turbulent, quadratic regime with stopping
times exceeding millions of years.

When the gas is more tenuous and optically thin, the smallest
particles dynamically damp, but are then driven away by radiation
pressure \citep{weiden1977a, ta2001}. This process plays a key role in
planet formation: growing planets stir smaller bodies, triggering a
collisional cascade that collisionally grinds debris into dust. The
tenuous gas disk and starlight remove material from the small-size tail of
the cascade. Thus, there is a mass ``drain'' for the system that can
limit the ultimate size of the planets that drive the cascade
\citep[see][]{knb2016}. This phenomenon is the basis for the present
work, in which the mass of Mars comes out small because of a
collisional cascade driven by dynamical shake-up.

\subsection{Secular evolution in the inner solar system}

Combining the physics of precession, eccentricity pumping and gas
damping, we can track orbits of planetesimals and protoplanets in the
inner solar system during the epoch of a sweeping $\nu_5$ resonance.
From Appendix~\ref{appx:secular}, the evolution equations for orbital
elements are
\begin{eqnarray}
\frac{de}{dt} & = & -(\ompres-\ompreJ)\eforce\sin(\varpi_s - \varpi_J)
+ e/\tauedamp,
\\
\frac{d\varpi}{dt} & = & 
(\ompres-\ompreJ)\left[1-\frac{\eforce}{e}\cos(\varpi_s - \varpi_J)
\right].
\end{eqnarray}
We integrate these equations to determine the orbital evolution of a
protoplanet, starting from a circular orbit at some early time. We
also specify an initial orbital configuration for Jupiter and Saturn,
as well as a decay mode for the disk. Our results were validated with
numerical simulations based on the $n$-body component of
\orchestra\ with a binned representation of the disk
(\S\ref{sssect:diskgrav}).

\begin{figure}[htb]
\centerline{\includegraphics[width=7.0in]{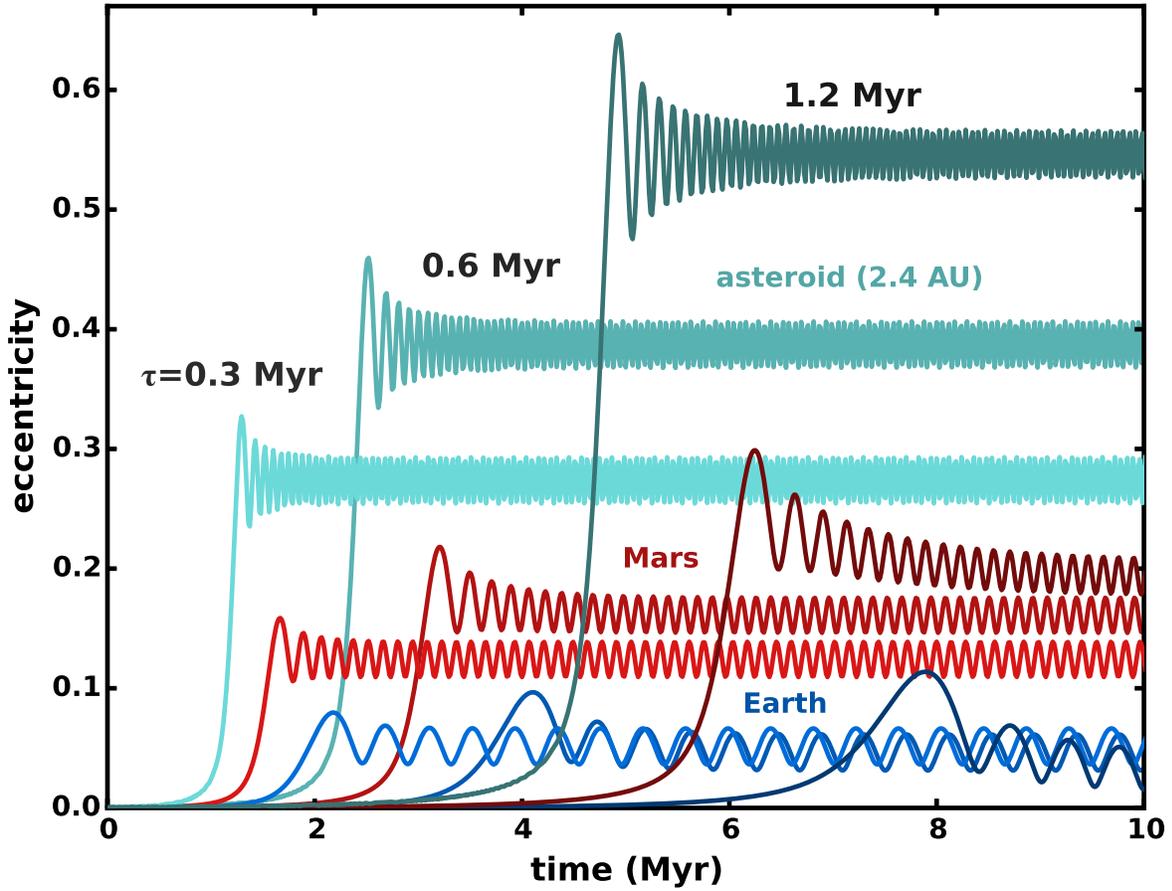}}
\caption{\label{fig:dedt} Eccentricity evolution in
  a gas disk that undergoes uniform depletion.  The plot shows the
  eccentricity of Mars (red curves) and a Ceres-size asteroid at
  2.4~AU (cyan curves) as a function of time. The three curves in each
  set are for different depletion time scales ($\tau$), as labeled
  (darker is slower depletion).  For Mars, some eccentricity damping 
  is apparent, but only for the more slowly decaying disk. 
%
%
  With a greater mass, the Earth damps more strongly than Mars and the 
  asteroid.  Note that the sweeping resonance hits earlier and with greater
  effect at larger orbital distance and when the decay of the disk is
  slow.  }
\end{figure}

\begin{figure}[htb]
\centerline{\includegraphics[width=7.0in]{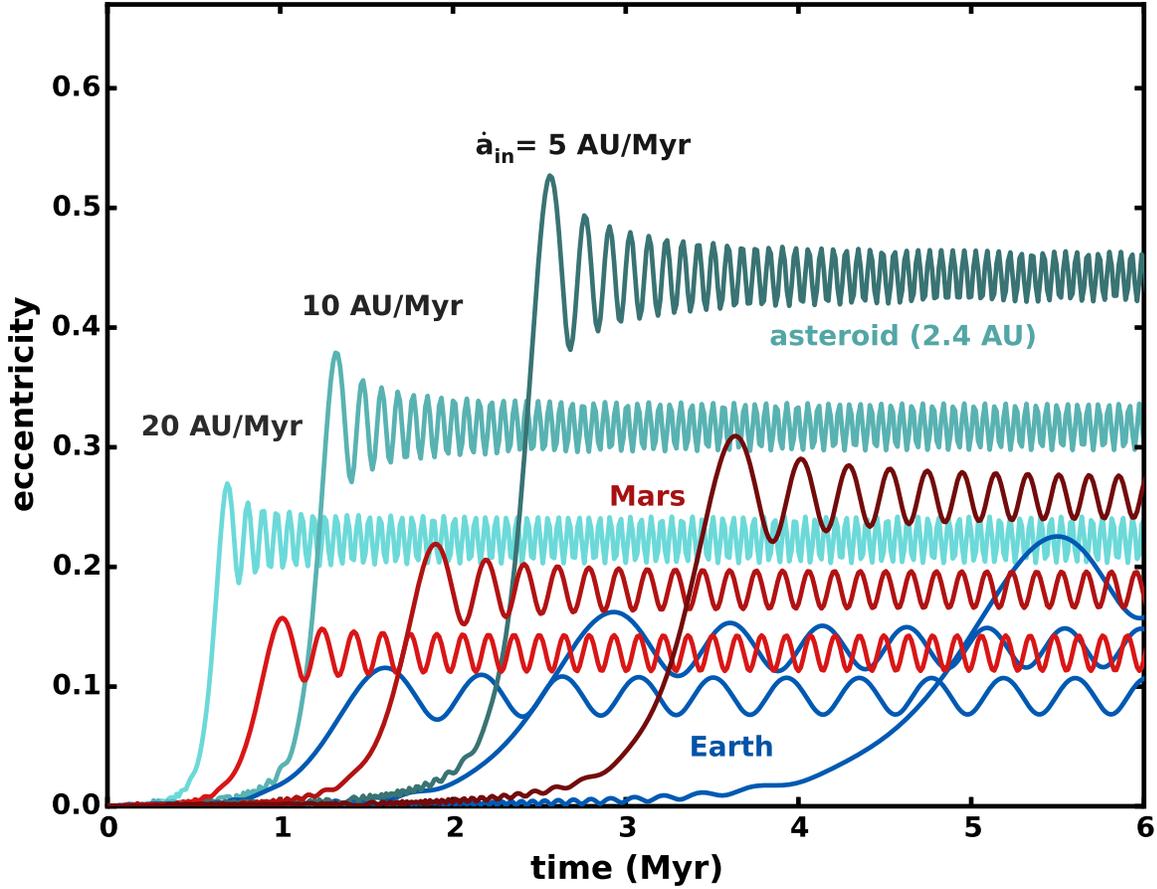}}
\caption{\label{fig:dedtinsideout} Eccentricity evolution in a
  disk with an expanding inner edge.  As in the previous Figure, the
  eccentricity of Mars is indicated by the red curves, with darker
  values corresponding to longer-lived disks (slower $\openrate$). The
  curves showing the evolution of a Ceres-like asteroid are in
  cyan. Slower disk evolution and closer proximity to Jupiter mean
  more eccentricity pumping.  Since the sweeping resonance hits when
  the disk's inner edge is well clear of the inner solar system, there
  is no gas drag in these models.  
}
\end{figure}

Figures~\ref{fig:dedt} and \ref{fig:dedtinsideout} show the secular
evolution of an Earth-mass planet at 1~AU, Mars at 1.52~AU, and a
Ceres-mass asteroid at 2.40~AU in response to Jupiter (5.20~AU) and
Saturn (9.58~AU). We assume that only Jupiter contributes to the
eccentricity pumping, and we adopt an eccentricity of $\ejup = 0.03$,
which gives an approximate time-average value as Jupiter's orbit evolves
under the influence of Saturn. To give a sense of the state of the
disk when the resonance sweeps inward, the surface density at Mars is
40~g/cm$^2$ (compared to an initial $\sim$1300~g/cm$^2$) in the
uniform depletion model.  In the inside-out erosion scenario, the
disk's inner edge crosses just beyond 15~AU when the resonance hits
Mars.

Figures~\ref{fig:dedt} and \ref{fig:dedtinsideout} illustrate that
proximity to Jupiter is a key factor in dynamical shake-up \citep[see
  also][]{naga2000, naga2005, thommes2008}. Planetesimals and
protoplanets closer to the gas giant experience more eccentricity
pumping. The rate of resonance sweeping is also important.  Rapidly
evolving disks (small $\tau$ or large $\openrate$) leave
protoplanetary orbits less excited. Comparing the final outcomes of
the uniform dissipation model and the inside-out clearing scenario, we
infer the role of eccentricity damping by gravity wakes
(Equation~(\ref{eq:edamp})); the uniform dissipation model has gas in
the inner solar system, so objects more massive than Mars orbitally
damp by this mechanism.  In all other cases involving these
short-lived gas disks, orbital damping by gravity wakes is not
important.

Although not included in the models described in this section, the
inner solar system may have had a long-lived residual disk,
replenished by comets or collisions between asteroids \citep[the disk
  around $\beta$~Pic provides an example;][]{lagrange1987, beust1996,
  czechowski2007, kral2016}.  Even with a low surface density ($\Sigma
\ll 10$~g/cm$^2$) this component may help to clear collisional debris
during planet formation \citep{knb2016} and damp eccentricities well
after the planets formed. From Equation~(\ref{eq:edamp}), we estimate
that a residual disk with a surface density of roughly 0.01~g/cm$^2$
can damp the orbit of Mars in roughly 1~Gyr.

\section{Numerical Simulations}
\label{sect:coag}

To illustrate the impact of dynamical shake-up on numerical
simulations of planet formation, we consider a representative example
designed to remove solid material at $a \gtrsim$ 1.5~AU well before
protoplanets reach the mass of Mars. From previous simulations of
terrestrial planet formation with little dynamical depletion at
1.5--3~AU \citep[e.g.,][]{kb2006, lunine2011, chambers2013, walsh2016,
  haghighipour2016}, giant impacts produce Mars-mass (Earth-mass)
objects in 1--10~Myr (10--100~Myr).  Radiometric analyses suggest Mars
achieved most of its final mass in $\lesssim$ 3--5~Myr
\citep{dauphas2011a,dauphas2011b}.  Thus, we focus on dynamical
shake-up scenarios where sweeping secular resonances pass through the
terrestrial zone at 1--2~Myr.

Within this scenario, we consider the following sequence of events.

\begin{description}

\item[Cloud collapse:] Current models envision formation of a star +
  disk system during the collapse of a dense core in a giant molecular
  cloud \citep[and references therein]{mckee2007}.  As cloud material
  falls onto the outer disk, the central protostar and inner disk
  eject material in a high velocity bipolar jet
  \citep[e.g.,][]{bontemps1996, bally2007}.  After roughly 0.5--1~Myr,
  infall and outflow effectively cease, leaving behind a fairly
  massive circumstellar disk surrounding a pre-main sequence star
  \citep[e.g.,][]{vankempen2009,eisner2012}.  Radiometric data and the
  demographics of circumstellar disks suggest formation of 10--100~km
  or larger planetesimals during this class I phase of evolution
  \citep{greaves2010b,dauphas2011a,najita2014}.  In our calculations,
  the formal $t$ = 0 occurs sometime during the class I phase.

\item[Gas giant formation:] Growth of gas giants is a several step
  process, involving (i) production of a multi-Earth-mass core of ice
  and rock, (ii) gradual accumulation of a gaseous atmosphere, and
  (iii) more rapid accretion of gas from the disk \citep{pollack1996,
    pierens2013, piso2014}.  We assume that Jupiter and Saturn grow
  quickly out of icy planetesimals on a time scale of $\tau_f \approx$
  0.5--1~Myr \citep{kb2009,bk2011a,levison2015,chambers2016}.  The
  nominal formation time is sensitive to the mode of disk dispersal:
  gas giants need to form more rapidly in disks dissipating
  from the inside-out than in disks depleting homologously with
  radius.  Depending on conditions within the disk, cores may migrate
  as they grow \citep[see, e.g.,][]{paarde2006,lyra2010,cossou2014,
    bitsch2015}.
  
\item[Disk dissipation:] Disk depletion occurs in parallel with the
  growth of gas giant planets. For simplicity, we adopt a uniform
  depletion model with $\tau = 0.3$~Myr.  Dynamical shake-up with
  inside-out depletion at $\openrate = 20$~AU/Myr has a similar impact
  on the growth of solids at 1.5--3~AU. The major difference between
  the two scenarios is the amount of residual gas available to clear
  collisional debris in the terrestrial zone.

\item[Gas giant migration:] Once gas giants reach a critical mass,
  they open up a gap and begin to migrate through the disk. When
  $\tau_f \ll \tau$, migration is ubiquitous
  \citep[e.g.,][]{walsh2011}.  If $\tau_f \gtrsim \tau$, however, the
  smaller surface densities in the disk lead to longer drift time
  scales (Equation~(\ref{eq:typeII})). Gas giants then may be unable
  to migrate far once they approach their final masses.

\item[Resonance sweeping:] As the disk depletes, resonant excitations
  sweep through the terrestrial zone. Material closest to the giant
  planets experience resonances before material closer to the Sun
  (Figure~\ref{fig:rezsweep}). In contrast, growth of solids is faster
  closer to the Sun. Thus, growth is limited most (least) severely in
  the asteroid belt (inside 1~AU). Depending on the level of gas
  depletion, aerodynamic drag may remove collision fragments with sizes
  of 1~cm and smaller \citep[see also][]{knb2016}.

\end{description}

In our approach, Jupiter and Saturn reach their final masses rapidly
and are in place when resonance sweeping begins. These assumptions are
reasonable for (i) systems with $\tau_f \ll \tau$, where the gas
giants finish any migration through the gas (including a Grand Tack)
prior to significant disk depletion and (ii) systems with $\tau_f
\approx \tau$, where the surface density of the disk is too small to
support much radial migration. We further assume that any migration of
the gas giants has a negligible impact on the surface density of
solids, which allows us to compare the depletion that might be
produced by migration \citep{walsh2011, batygin2015} with the
depletion due to resonance sweeping.

To follow the evolution of solids in the terrestrial zone, we perform
a coagulation calculation \citep{saf1969, spaute1991, kl1998, oht2002,
  kb2006, kb2008}.  In our approach, described in
Appendix~\ref{appx:numsim}, we seed 32 concentric annuli with a swarm
of 1--100~km planetesimals having $\Sigma_s \propto a^{-3/2}$ over
1--3~AU. Within each annulus, we track the mass and velocity evolution
of particles with sizes ranging from a minimum size \rmin\ to a
maximum size \rmax. For improved accuracy, the mass spacing factor
between adjacent bins is $\delta = 2^{1/4}$.  Using well-tested
statistical techniques, the code calculates the outcomes of
gravitational scattering and physical collisions among all mass bins
in all annuli. The appendix describes the algorithms in more detail.

For this study, we add a mechanism for eccentricity pumping from
Jupiter during the phase of the sweeping secular resonance:
\begin{equation}
\dot{e} = \begin{cases}
 \frac{1}{\sqrt{2}} \epumprate
     & \ \ \ \textrm{if $(e < \sqrt{2}\eforce)$}
\\ 0 & \ \ \ \textrm{otherwise.}
\end{cases}
\end{equation}
This algorithm works well in reproducing the eccentricity pumping in
numerical experiments using secular perturbation theory and the
$n$-body component of \orchestra. In the presence of a dissipating gas
disk, we apply this algorithm at all times. If the disk is static or
absent altogether, we turn off this stirring.

For this calculation, we consider a mono-disperse set of solid planetesimals 
with initial radius \r0\ = 100~km and total mass \M0.  The planetesimals have 
mass density \rhop\ = 3~\gcmc, initial eccentricity $e_0 = 10^{-3}$, and 
initial inclination $\imath_0 = e/2$.  We consider grids extending from 1~AU 
to 3~AU with \rmin\ = 1~\mum\ or 1~cm. For simplicity, we neglect gas drag. 
Throughout the evolution we consider, collisional damping, dynamical
friction, and viscous stirring dominate reduction of small particle velocities
by gas drag \citep[e.g.,][]{weth1993,kl1998}. However, the gas can still 
generate a substantial radial drift, which removes small particles from the 
grid \citep[e.g.,][]{knb2016}. Adopting two different values for \rmin\ 
allows us to quantify the potential importance of radial drift without 
spending extra cpu time on a more complicated calculation. Here our goal 
is to learn whether dynamical shake-up can prevent the growth of protoplanets 
inside of 3~AU.  In a more realistic system, gas drag probably aids planet 
growth at the inner edge of the grid and prevents growth at the outer edge.

\section{Results}
\label{sect:num}

In calculations with no sweeping resonance, growth follows a standard
pattern.  Collisions among 100~km planetesimals with $e_0 = 10^{-3}$
produce mergers and some debris. As the largest objects grow,
dynamical friction damps their orbits. Damping leads to larger
gravitational focusing factors and a short phase of runaway growth
when protoplanets reach radii of 1000-2000~km. During runaway growth,
viscous stirring excites the orbits of leftover planetesimals and
smaller particles of debris. Stirring initiates a collisional cascade
which grinds leftovers into smaller and smaller particles that are
ejected by radiation pressure. Loss of material slows the growth of
protoplanets, which reach a characteristic maximum size of 3000~km.

When the resonance sweeps through the grid, it tends to drive all of
the solids to large $e$ (e.g., Figure~\ref{fig:dedt}). Among the
solids, dynamical friction tries to damp $e$ and $i$ of the largest
protoplanets. When the resonance is strong (weak), dynamical friction
cannot (can) keep up with resonant stirring. Thus, strong resonances
drive a collisional cascade before protoplanets have a chance to grow.

\begin{figure}[htb]
\centerline{\includegraphics[width=7.0in]{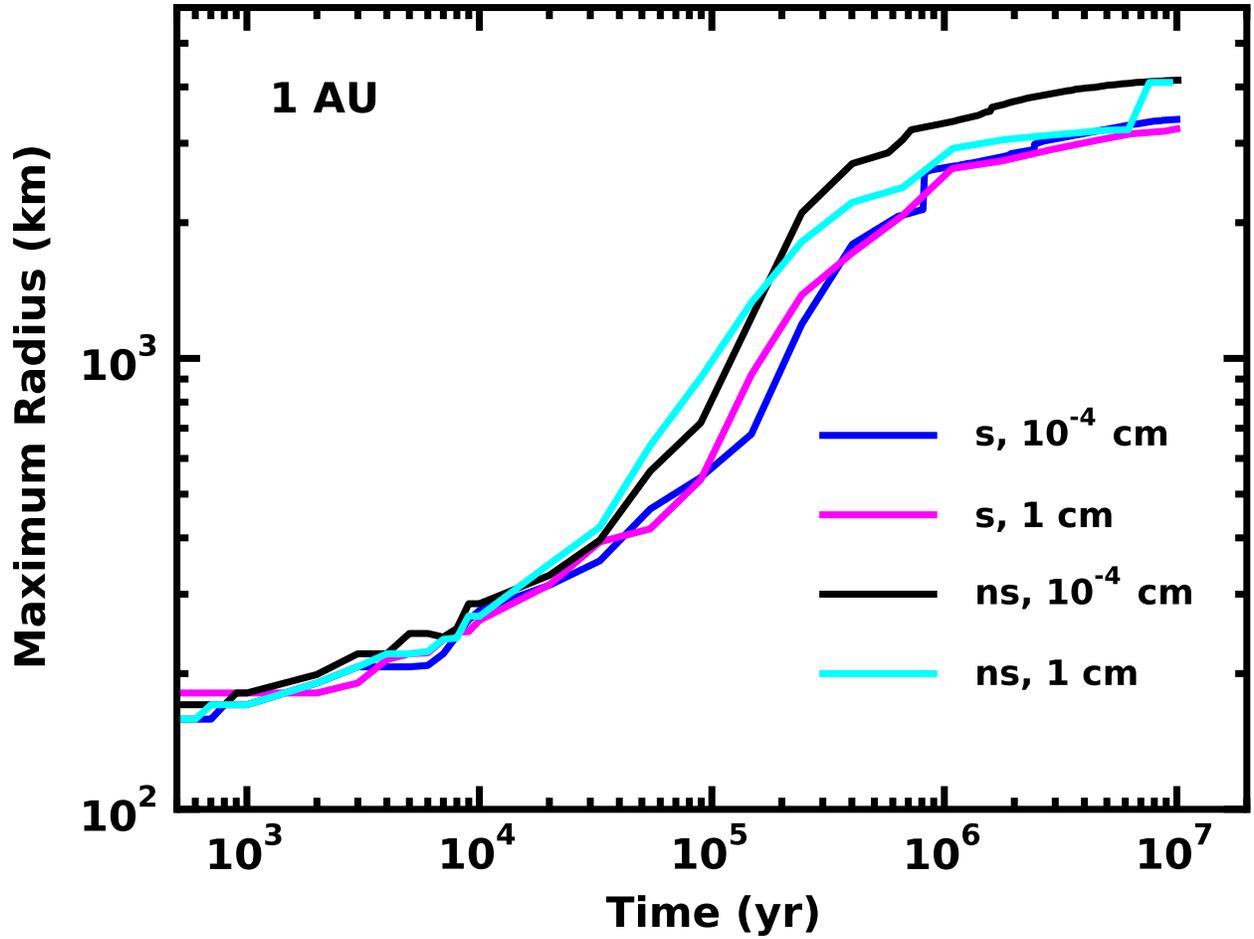}}
\caption{\label{fig:rmax1au} Time evolution of \rmax\ the size of the
  largest object at 1~AU for systems starting with 100~km
  planetesimals, \rmin\ = 1~\mum\ or 1~cm (as labeled), and with (`s')
  or without (`ns') stirring from dynamical shake-up.  The magnitude
  of dynamical shake-up and the size of the smallest object in the
  coagulation grid have little impact on collision outcomes at 1 AU.
}
\end{figure}

Figure~\ref{fig:rmax1au} illustrates the evolution of the largest
objects at 1 AU in four separate simulations. Early on, 100~km objects
collide and merge into larger objects. After $1-2 \times 10^4$~yr, the
largest objects reach 300~km sizes. By $10^5$ yr, mergers have
produced a set of 500--600~km objects. After $10^5$ yr, the largest
objects in the simulations follow similar but somewhat divergent
tracks.  Fairly independent of \rmin, systems with no dynamical
stirring (labeled `ns' in the Figure) grow somewhat faster than
systems with dynamical stirring (labeled `s').  After 10~Myr, the
largest objects in the `ns' (`s') tracks reach sizes of 4000~km
(3000~km).

\begin{figure}[htb]
\centerline{\includegraphics[width=7.0in]{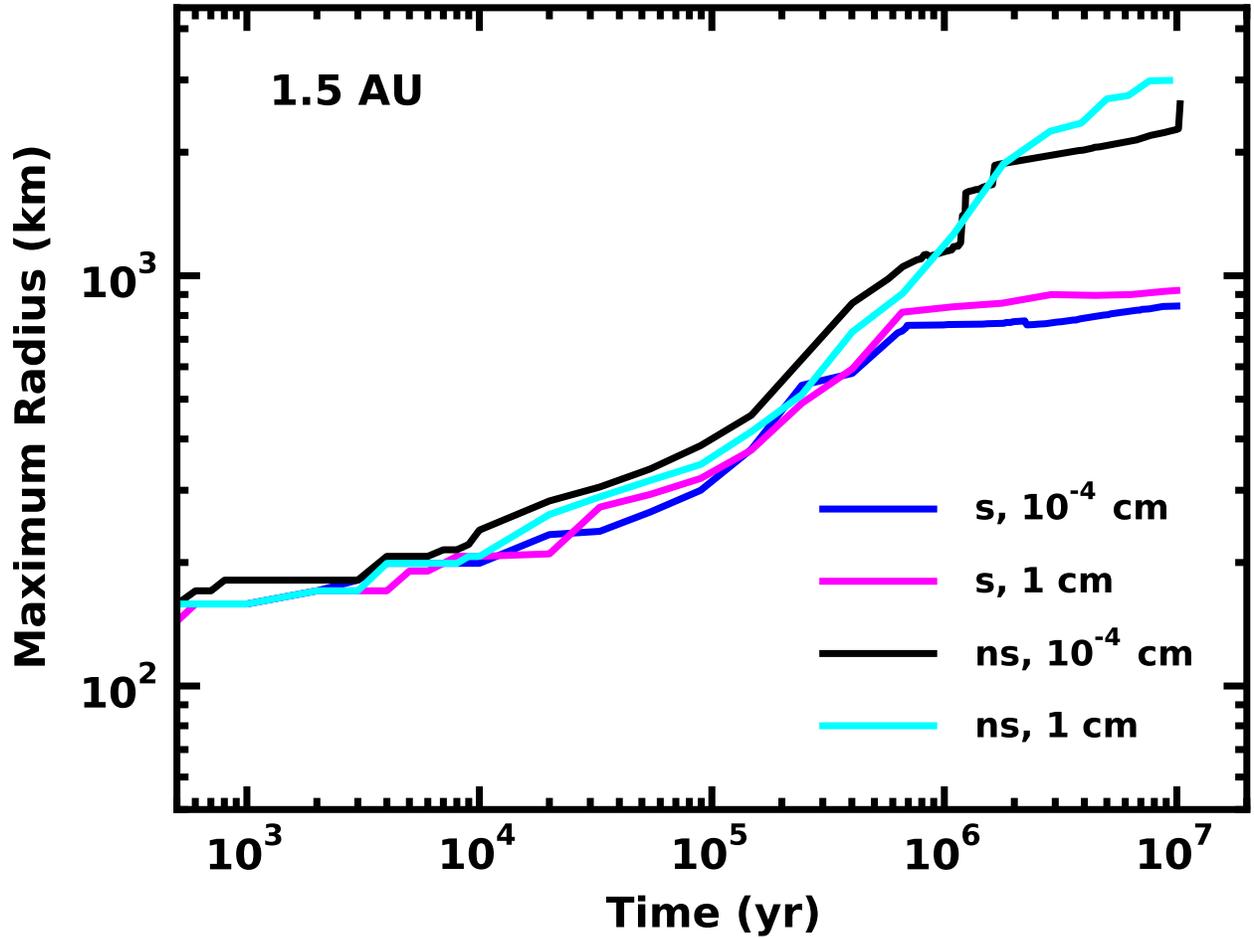}}
\caption{\label{fig:rmax1p5au} As in Figure~\ref{fig:rmax1au} for
  calculations at 1.5~AU. Collision outcomes do not depend on \rmin
  (values are listed in the legend); however, dynamical shake-up ('s')
  reduces \rmax\ by factors of 3--5 on time scales of 1--10~Myr as compared with
  no dynamical stirring ('ns').  }
\end{figure}

At 1.5~AU, the differences in the evolution are more dramatic
(Figure~\ref{fig:rmax1p5au}).  As in systems of planetesimals at 1~AU,
the evolution for 0.1--0.3~Myr is independent of \rmin\ or the level
of stirring. By 1~Myr, slower growth in systems with dynamical
stirring is evident. After 10~Myr, systems with (without) stirring
have objects with maximum sizes of 800--1000~km
(2000--3000~km). Independent of stirring, the maximum size is
insensitive to \rmin.

\begin{figure}[htb]
\centerline{\includegraphics[width=7.0in]{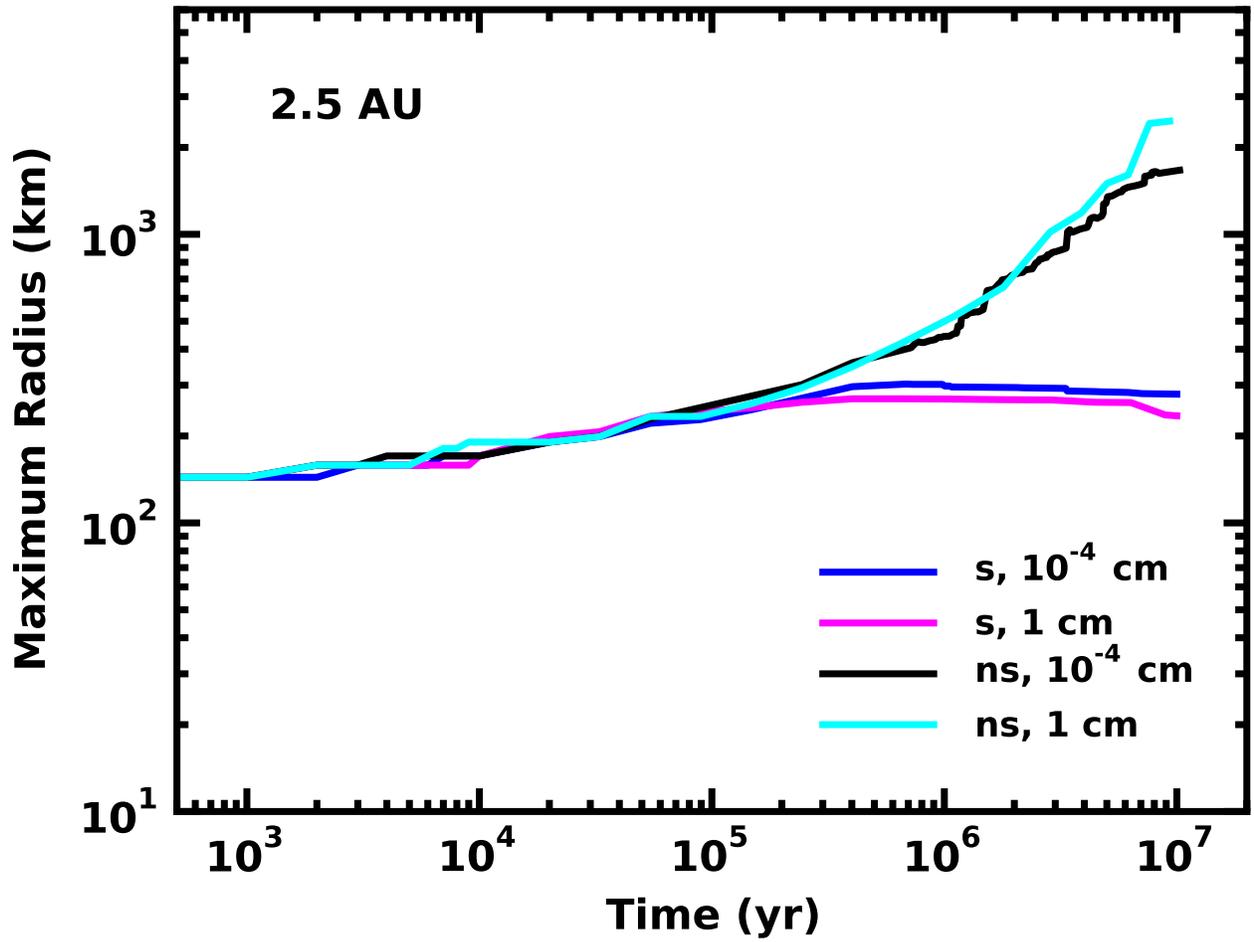}}
\caption{\label{fig:rmax2p5au} As in Figure~\ref{fig:rmax1au} for
  calculations at 2.5~AU.  Dynamical shake-up ('s'; \rmin\ values are
  shown) reduces \rmax\ by factors of 10, compared to calculations
  with no stirring ('ns').  }
\end{figure}

At 2.5~AU, dynamical stirring prevents large objects from growing past
200--300~km (Figure~\ref{fig:rmax2p5au}). After 0.1--0.3~Myr of
growth, the largest objects reach typical maximum sizes of 300~km. In
models with no dynamical stirring, the largest objects grow slowly to
500~km sizes at 1~Myr and 2000~km sizes at 10~Myr. In models with
dynamical stirring, however, growth effectively ceases once stirring
begins.  As the evolution proceeds, high velocity collisions with
small objects gradually diminish the sizes of the largest
objects. After reaching peak sizes of 300~km at roughly 1~Myr, the
largest objects have radii of roughly 200~km after 10--100~Myr.

In this example, collisional cascades driven solely by dynamical
shake-up remove a large fraction of solid material from the
system. When \rmin\ = 1~\mum\ (1~cm), systems with dynamical shake-up
lose 30\% (95\%) of their initial mass in solids.  In models without
dynamical shake-up, mass loss by collisional disruption is much less
severe, $\lesssim$ 5\% when \rmin\ = 1~\mum\ and 60\% when \rmin\ =
1~cm.

These results demonstrate that sweeping secular resonances have a
profound influence on the outcomes of terrestrial planet formation
inside the orbit of Jupiter \citep[see also][]{thommes2008}.  For the
initial conditions examined here, there is little impact on planet
formation inside 1~AU: planet formation is fairly rapid and dynamical
shake-up is rather weak.  At larger $a$ where the resonance is much
stronger, the growth of planets is much slower.  At 1.5~AU,
protoplanet growth stalls at 1000~km. At 2.5~AU, protoplanets barely
grow larger than 300~km.

The models considered here assume that Saturn orbits at its current
position.  Unless Saturn is closer than the 3:2 orbital
commensurability with Jupiter, relaxing this assumption does not
change our results significantly.  In compact configurations, the
$\nu_5$ resonance crosses just inside of Mars' orbit and fails to
reach 1~AU (Fig.~\ref{fig:rezsweepnewsaturn}). Dynamical excitation
may then completely clear a narrow region between Earth and Mars,
without stirring Earth's progenitors at all.  Moving both gas giants
to larger orbital distance, decreasing their separation, or both, can
prevent the sweeping resonance from reaching Mars.  Moving both giants
closer can destructively stir material even at Earth's location.
Thus, the orbits and masses of Earth and Mars provide constraints on
the history gas giant's orbital configuration
\citep[see][]{brasser2009, minton2011, agnor2012, kaib2016,
  haghighipour2016}.

\section{Implications for exoplanetary systems}

Dynamical shake-up is likely to occur around stars other than the
Sun. We expect sweeping resonances within circumstellar disks in
stellar binaries \citep[e.g.,][]{hepp1978, kley2008, paardekooper2008,
  raf2013} and in circumbinary disks \citep[e.g.,][]{marzari2008,
  silsbee2015a}.  For star-planet binaries like the Sun and Jupiter,
and for regions interior to the giant planet, we can use
Equation~(\ref{eq:epump}) for eccentricity pumping to estimate under
what conditions a dynamical shake-up is important. By multiplying the
pumping rate and the gas disk life-time, we find
\begin{equation}\label{eq:shakeup}
e_{\rm bin} \, \mu_{\rm bin} \,
\left(\frac{T_{\rm gas}}{T_{\rm bin}}\right)
\left(\frac{T}{T_{\rm bin}}\right)^{\!5/3}
\ \gtrsim \ 0.1\,,
\end{equation}
where $e_{\rm bin}$ is the binary eccentricity, $\mu_{\rm bin}$ is the
giant planet's mass relative to the host star, $T_{\rm bin}$ is its
orbital period, $T_{\rm gas}$ characterizes the gas disk life time,
and $T$ is an interior body's Keplerian orbital period. For
Earth's location in the Sun-Jupiter binary, and with $T_{\rm gas}$ set to
$10^6$~yr, the left-hand side of the expression is roughly 0.05, suggesting
that dynamical shake-up is not important. On the other hand, the left-hand
side exceeds the threshold value of 0.1 for both Mars at 1.5~AU and
an asteroid at 2.5~AU, with values of 0.13 and 0.45, respectively. 

Stars with a massive planet near or beyond the snow line are fairly
common. Radial velocity surveys \citep{bonfils2011} and microlensing
studies \citep{shvartzvald2016} suggest that as many as 50\% of M
dwarfs host icy super-Earths, and up to 5\% host Jupiter-mass
planets. Eccentricity estimates of all known long-period exoplanets
are consistent with $e \gtrsim 0.02$ \citep{exoplanets2014}.
Equation~(\ref{eq:shakeup}) suggests that even super-Earths can cause
a shake-up, when (i) the mass of the host star is low, (ii) the gas
dissipation time scale is long, and/or (iii) the eccentricity of the
massive planet is large. Thus, scaled-down versions of asteroid belts
around M dwarfs are plausible.

Unless the star's disk dissipates slowly or the planet's orbit is 
strongly eccentric, a Sun-like star requires a Jupiter-mass planet 
for shake-up.  Radial velocity studies suggest that 5--10\% of Sun-like
stars host gas giants out to 5~AU \citep{cum2008, zec2013}.  Consistent 
with theory \citep{kk2008}, more massive (A- and F-type) stars show a 
slight increase in the frequency of giant planets 
\citep{lovis2007, johnson2007, bowler2010}.  While most of the observed
giant planets have shorter orbital periods than the
Sun-Jupiter binary, they still admit the possibility of an extended
region in the terrestrial zone that experienced dynamical shake-up. We
conclude that asteroid belt analogs may be found interior to the gas
giants in a broad range of stellar hosts.

From our sample calculations of dynamical shake-up, planetary systems 
that experience a sweeping resonance have an early, enhanced production 
of dusty debris compared to systems with no outer gas giant. In systems
where the gas has negligible surface density, copious dust production 
is probably visible \citep[e.g.,][]{kb2004b,kb2016a}. If the disk has 
a modest surface density of 0.01--20~g~cm$^{-2}$ during the shake-up, 
however, collisional debris is probably cleared out rapidly by gas drag
and radiation pressure \citep{weiden1977a, ta2001}. In these systems,
rocky planet formation may be ``quick'' and ``neat'' \citep{knb2016}.

\section{Summary}
\label{sect:done}

In this paper, we describe a scenario for rocky planet formation that
includes a sweeping secular resonance driven by Jupiter and the
dissipating solar nebula. As in the dynamical shake-up models proposed 
in previous work \citep{naga2005, nagasawa2007, obrien2007, thommes2008}, 
we try to explain the low mass of Mars and the asteroid belt.  Our new 
contribution is to explore outcomes when the resonance sweeps
through the inner solar system quickly, before the formation of Mars
is complete. This key modification allows Jupiter's eccentricity pumping 
to increase fragmentation from high velocity collisions during the runaway 
and oligarchic phases of planet formation. Coupled with clearing of small 
debris particles by gas drag and radiation pressure, this process yields 
a way to inhibit the growth of protoplanets beyond 1~AU.

Our coagulation calculations with the \orchestra\ code illustrate that
an early sweep of the $\nu_5$ secular resonance explains the low mass
of Mars, as well as the sizes of objects and total mass in the asteroid 
belt. We ran two extensive multi-annulus coagulation simulations, with 
and without a sweeping resonance.  As in other simulations of planet 
formation within the terrestrial zone \citep[e.g.,][]{chambers2001a, 
obrien:2006b, kb2006, raymond2009, morishima2010}, calculations without the 
resonance yielded Earth-size protoplanets on a Myr time scale.  In 
contrast, calculations with the sweeping resonance produce protoplanets 
that reach only $10^3$~km in a significantly depleted disk.

Models with a dynamical shake-up at later times, with disk dissipation
time scales of 5--10~Myr, successfully limit the mass of Mars and
clear the asteroid belt \citep[e.g.,][]{nagasawa2007}.  In these
scenarios, orbital energy losses of solids in a more slowly sweeping
resonance and a longer-lived gas disk become important, and material
drifts radially inward.  The building blocks of the rocky planets then
originate as far way as the asteroid belt and migrate inward
with the resonance.  The formation time for Mars in these scenarios
exceeds 10~Myr, consistent with available radiometric data at the time
\citep{jacobsen2005}. Our version of the shake-up scenario with {\it
  in situ} formation is motivated by more recent evidence for a
shorter formation time for Mars \citep{dauphas2011b}.

In dynamical shake-up scenarios like those discussed here, Jupiter is
in place near its present orbital distance. While we also choose
Saturn's orbit to be similar to its present-day configuration, a
shake-up will occur at Mars' location and in the asteroid belt even if
Saturn is near the 3:2 commensurability, around 7~AU.  Thus, dynamical
shake-up may occur as a precursor to late-time orbital instability
scenarios like the Nice model \citep{tsig2005}, which requires
Saturn's to orbit closer to Jupiter than it is now.

Jupiter's orbital eccentricity is key to dynamical shake-up.  If the
gas giant was assembled close to its current location, as in
``classical'' models \citep[see][and references
  therein]{haghighipour2016}, interactions with the gas disk produce a
large enough eccentricity to drive a shake-up \citep{gold2003,
  duffell2015}. In other scenarios, convergent migration leads to
resonance trapping between Jupiter and Saturn \citep{masset2001,
  morbycrida2007}. The Grand Tack \citep{walsh2011} is one example,
tuned to trigger the resonance trap and mutual outward drift only
after the gas giants are deep inside the terrestrial zone.  Jupiter's
eccentricity in these models may have been low
\citep[e.g.,][]{morbytsig2007, deienno2016}.  However, 2D
hydrodynamical calculations suggest that with Jupiter and Saturn in a
2:1 mean-motion resonance \citep[a configuration favored in the Nice
  model;][]{nesvorny2012}, Jupiter's eccentricity is $\ejup \approx
0.03$ \citep{pierens2014}, as in our shake-up models. There is also
evidence from 3D simulations that $\ejup$ can grow to this level when
Jupiter and Saturn occupy the 3:2 resonance \citep{dangelo2012}. Thus,
conditions for dynamical shake-up arise in a variety of scenarios.

As a constraint for planet formation theory, the mass of Mars is a
challenge \citep{weth1991}.  Models allow a range of possibilities,
including an initially truncated protoplanetary disk at 1~AU
\citep{jin2008, hansen2009, izidoro2014, haghighipour2016}, dynamical
excitation and depletion \citep{weth1992, raymond2009}, and the Grand
Tack \citep{walsh2011}.  Compared to these scenarios, a dynamical
shake-up has a clear advantage: a sweeping resonance generated by the
dissipation of a disk where the gas giants formed seems inevitable.
Our work here suggests that this phenomenon alone can explain the low
mass of Mars and a depleted asteroid belt.

\textbf{
As these theoretical scenarios become more complete, it should become
possible to predict the formation time scale of Mars, $\tau_f$, as
well as the dynamical architecture of Mars and the asteroid belt as
functions of $\tau_f$ and the gas disk lifetime, $\tau$.  We expect
that systems with $\tau_f \ll \tau$ will be prone to depletion from
Grand Tack {\it and} dynamical shake-up; systems with longer $\tau_f$
may only undergo dynamical shake-up. If this hypothesis is correct,
disks with longer $\tau_f$ may yield more massive analogs of Mars and
the asteroid belt than those with smaller $\tau_f$. 
%
%
As observational constraints on $\tau$ and $\tau_f$ improve 
\citep[e.g.,][]{dauphas2011a, connelly2012, tang2014, morris2015,
schiller2015, dauphas2017, fischer2017, wang2017}, it will be possible
to test this prediction.
}
%

Our models have several connections to observations of exoplanetary
systems.  If dynamical shake-up is inevitable for a planetary system
with a long-period giant planet, asteroid belt analogs should be as
common as gas giant planets.  Roughly 10\% of Sun-like stars with
Jupiter-mass planets and a higher fraction of M dwarfs with
super-Earths and Neptunes are configured in this way. Assuming an
initially massive disk of solids between the super-Earth/gas giant and
the central star, Mars analogs and extrasolar asteroid belts should be
similarly frequent. Future exoplanet searches may detect these planets
directly from transits \citep[e.g.,][]{borucki2013}, while sensitive
microlensing surveys may reveal extrasolar asteroid belts
\citep{lake2017}.

Observations of young planetary systems may reveal dynamical shake-up
in action, since a large amount of debris is produced compared to
systems without a shake-up. Detection of this debris depends on the
timing of resonance sweeping and the manner of disk depletion. Unless
the gas drags solid material into the central star \citep{knb2016}, we
expect brighter dust signatures in transition disk systems with gas
giant planets at 5--10 AU than in those without gas giants.

The results of this preliminary study are promising. In future work we
will include a more accurate treatment of the dynamics within
\orchestra's \nbody\ code. Then we can begin to compare the final
outcomes of an early dynamical shake-up with the observed orbits of
the Sun's asteroids. We also plan to include more detailed modeling of
the small debris from collisions. Our assumptions here are that this
debris is rapidly cleared from the terrestrial zone, making its
detection in this region difficult \citep[cf.][]{kb2004b}. If this
idea is true, it would explain the low incidence of debris disks in
the terrestrial zones of other stars \citep[e.g.,][]{mamajek2004}
while still offering hope that these stars host planets like Earth.

\acknowledgements

We are grateful to NASA for a generous allocation of supercomputing
time on the 'discover' cluster.  This project was supported in part by
the NASA Outer Planets Program through grant NNX11AM37G.  This
research has made use of the Exoplanet Orbit Database and the
Exoplanet Data Explorer at exoplanets.org.

\appendix

\section{Theory of circumstellar orbits in a binary system}
\label{appx:mostcirc}

Here we derive analytical expressions that describe the orbit of a
small body, such as an isolated planetesimal or protoplanet, in a
binary system.  Taking the approach of \citet{lee2006} and
\citet{leung2013} for circumbinary orbits, our starting point is the
restricted three-body problem \citep[see][]{szeb1967,murray1999}.  The
primary and secondary have masses $\Mp$ and $\Ms$, respectively; the
binary has separation $\abin$ and eccentricity $\ebin$. To quantify
positions and velocities, we choose a cylindrical coordinate system
with the binary at zero altitude ($z=0$) and an origin located on the
primary or the binary's center of mass, depending on whether we are
interested in circumbinary (P-type) or ``\circpri'' (S-type) orbits.
In the \circpri\ case, we take the ``primary'' to mean the star that
hosts the planetesimal or protoplanet in question.  The theory
presented below holds even if the ``secondary'' is more massive than
the primary.

The gravitational potential at the position $(R,\phi,z)$ of a small
body on an \circpri\ orbit is
\begin{equation}
\label{eq:Phi}
\Phi = -\frac{G \Mp}{\sqrt{R^2+z^2+\Rp^2+2R \Rp \cos\Delta\phi}}
- \frac{G \Ms}{\sqrt{R^2+z^2+\Rs^2-2R \Rs \cos\Delta\phi}},
\end{equation} 
where $\Rp$ ($\Rs$) is the distance from the origin to the primary
(secondary), $G$ is the gravitational constant, and $\Delta\phi$ is
the small body's position angle relative to the direction of the
secondary. The gradient of this potential contributes to the
small object's equations of motion, from which we obtain orbit solutions
by following \citet{lee2006} and \citet{leung2013}: We first express the
potential as a cosine series, taken to first order in the binary
eccentricity. Then, we adopt a coordinate system that describes the
small body's position and velocity relative to a circular guiding
center, and linearize the equations of motion in these
coordinates. This strategy allow orbits to be solved in the manner of
a driven harmonic oscillator. \citet{lee2006} and \citet{leung2013}
provide details, along the lines of these instructions:

\begin{enumerate}
\item Create a series expansion of the potential,
first in terms of powers of angle cosines, then rewritten with harmonics: 
\begin{equation}\label{eq:Phiseries}
\Phi = \sum_{k=0}^\infty P_k \cos^k(\Delta \phi)
 = \sum_{k=0}^\infty P^\prime_k \cos(k\Delta \phi),
\end{equation}
using multiple angle formulae. For the \circpri\ case, we set the
coordinate origin to be coincident with the primary, and we modify
the potential to account for this choice of a non-inertial frame:
\begin{equation}
\Phi \rightarrow \Phi + \frac{G\Ms R}{\Rs^2}\cos(\Delta \phi).
\end{equation}
The new term compensates for the motion of the primary
in the equations of motion.

\item Include the dependence on binary eccentricity by writing
\begin{equation}\label{eq:rebin}
\Rs/\abin \approx 1-\ebin\cos\nbin t,
\end{equation}
and
\begin{equation}
\Delta \phi = \phi - \phis \approx \nbin t + 2\ebin\sin \nbin t
\end{equation}
where $\phis$ is the azimuthal coordinate of the secondary, $\nbin$ is
the binary's mean motion, and time $t$ is chosen to place the binary 
at periapse when $t=0$ and $\phis = 0$. In the circumbinary case,
Equation~(\ref{eq:rebin}), along with one like it for $\Rp$, 
includes a mass ratio factor to account for center-of-mass coordinates
\citep[see][]{leung2013}.

\item
Expand the potential to first order in $\ebin$ with the form
\begin{eqnarray}
\Phi(z=0) & = & \sum_{k=0}^\infty\left[ \Phi_k - \ebin \cos(\nbin t) \Phi^e_k
\right] \cos(k\Delta \phi)
\\
& = & \sum_{k=0}^\infty\Bigg\{ \Phi_k \cos k(\phi-\nbin t) + \\
\nonumber
&  &  
\ebin \bigg[
\left(k\Phi_k-\onehalf\Phi^e_k \right)\cos (k\phi-(k+1)\nbin t) + \\
\nonumber
&  & \ \ \ \ \,  
\left(-k\Phi_k-\onehalf\Phi^e_k \right)\cos(k\phi-(k-1)\nbin t) 
\bigg]
\Bigg\}
\end{eqnarray}
where the upper equation defines $\Phi_k$ and $\Phi^e_k$ (following
the convention of \citealt{leung2013}) as expansion coefficients when
we take into account the dependence on $\ebin$ in the binary's radial
position ($\Rp$ and $\Rs$, as in Equation~(\ref{eq:rebin})) but not in 
the angular coordinates $k\Delta\phi$.  The lower equation, which shows 
full first-order dependence on $\ebin$, takes advantage of multiple angle 
formulae and a Taylor expansion of $\cos\Delta\phi$ to express the potential 
as a cosine series.  Here we limit our analysis to the plane of the binary.
%

\item
Solve equations of motion, including
\begin{equation}
\ddot{R}-R\dot{\phi}^2 = -\frac{\partial \Phi}{\partial R}
\ \ \ \ \ \textrm{and} \ \ \ \ \ 
R\ddot{\phi}+2\dot{R}\dot{\phi} = 
    -\frac{1}{R}\frac{\partial \Phi}{\partial \phi},
\end{equation}
in terms of variables $\rgc + R_1$, and $\ngc t +
\phi_1$, where $\rgc$ is the orbital radius of a guiding center on a
circular path about the primary (or center of mass, in the
circumbinary case), and $\ngc$ is the mean motion at that orbital
distance, given by
\begin{equation}
\label{eq:ngc}
\ngc^2  = \frac{1}{\rgc} \left.\frac{d\Phi_{0}}{dR}\right|_\rgc.
\end{equation}
Keeping only terms that are linear in variables $R_1$ and $\phi_1$,
use the approximation of $\Phi$ as a sum of harmonic modes. The
equations of motion and their solutions are the same as in a simple,
driven harmonic oscillator; solutions to $R$, $\phi$ (and $z$, it
turns out) will also be sums of these same modes.

\end{enumerate}
These steps, with straightforward modification to include motion
in the $z$ direction (out of the binary's orbital plane), lead to
these solutions:
\begin{eqnarray}
\label{eq:mostcircr}
  R(t)& = &\rgc\left\{1 - \efree\cos(\kappae t+\psi_e) 
 \ -  
     \sum_{k=1}^\infty {C_k}\cos(k(\dn + \pagcb)t)
\right.
\\ \nonumber
& & 
\left.
- \ebin \left[
\myC^e_0\cos(\nbin t + \psibin) +
  \sum_{k=1}^\infty \myC^+_k \cos(k(\dn-\nbin+\pagcb)t-\psibin)
\right.\right. \\ \nonumber & & \left.\left. \ \ \ \ \ \ \ \ \ \ \ \ \ \ \ \ \ \
\ \ \ \ \ \ \ \ \ \ \ \ \ \ \ \ \ +
 \sum_{k=1}^\infty 
  \myC^-_k \cos(k(\dn+\nbin+\pagcb)t+\psibin)
  \right]\right\}
\\ 
\label{eq:mostcircphi}
  \phi(t) & = &\ngc\left\{ t + \frac{2\efree}{\kappae} 
    \sin(\kappae t+\psi_e) \ \ +   
     \sum_{k=1}^\infty \frac{D_k}{k\dn}\sin(k(\dn+\pagcb)t) \right.
\\
& & \nonumber
\left. + \ebin \left[\frac{\myD^e_0}{\nbin}\sin(\nbin t+\psibin) +  
    \sum_{k=1}^\infty 
\frac{\myD^+_k \sin(k(\dn-\nbin+\pagcb)t-\psibin)}{k\dn-\nbin}
\right.\right. \\ \nonumber & & \left.\left. \ \ \ \ \ \ \ \ \ \ \ \ \ \ \ \ \ \
\ \ \ \ \ \ \ \ \ \ \ \ \ \ \ \ \ \ \  +
 \sum_{k=1}^\infty 
        \frac{\myD^-_k \sin(k(\dn+\nbin+\pagcb)t+\psibin)}{k\dn+\nbin}
    \right]\right\}
\end{eqnarray}
\begin{eqnarray}
\label{eq:mostcircz}
& &
z(t)  =   i \rgc \cos(\kappai t + \psi_i), 
\hspace{3.75in}
\end{eqnarray} 
where $\dn \equiv \ngc-\nbin$ (a negative value indicates a
circumbinary orbit), $\efree$ is the ``free'' eccentricity, $i$ is the
inclination, $\pagcb \equiv \psigc-\psibin-\varpibin$, where
$\varpibin$ is the binary's longitude of periastron, and the phase
angles $\psigc$, $\psibin$, $\psi_e$ and $\psi_i$ are
constants. The epicyclic frequencies that appear in these solutions
are
\begin{equation}\label{eq:kappas}
\kappae^2 \equiv \rgc \left.\frac{d\ngc^2}{dR}\right|_\rgc\!\! + 4\ngc^2
\ \ \ \ \ \textrm{and} \ \ \ \ \ 
\kappai^2 \equiv \left. \frac{1}{z}\frac{d\Phi}{dz}\right|_{z=0,\rgc},
\end{equation}
while the mode amplitudes are
\begin{eqnarray}
  \label{eq:Ck}
C_k &  = & \frac{1}{\rgc(\kappae^2-k^2\dn^2)}
\left[\frac{d\Phi_k}{dR}+\frac{2\ngc\Phi_k}{\rgc\dn}\right]_{\rgc},
\\
\myC^e_0 & = & -\frac{1}{\rgc(\kappae^2-\nbin^2)}
\left[\frac{d\Phi^e_0}{dR}\right]_{\rgc},
\\
\myC^\pm_k & = & 
\frac{1}{R[\kappae^2-(k\dn\pm\nbin)^2]} 
\left[\pm k\frac{d\Phi_k}{dR}-\frac{1}{2}\frac{d\Phi^e_k}{dR} +
\frac{k\ngc(\pm 2k\Phi_k - \Phi^e_k)}{R(k\dn\pm\nbin)}\right]_{\rgc},
\end{eqnarray}
and
\begin{eqnarray}
  \label{eq:Dk}
D_k & = & 2C_k - \left[\frac{\Phi_k}{\rgc^2\ngc\dn}\right]_{\rgc},
\\
 \myD^e_0 & = & 2\myC^e_0, 
\\[4pt]
\myD^\pm_k & = & 2\myC^\pm_k 
- \left[\frac{k(\pm2k\Phi_k-\Phi^e_k)}{2\rgc^2\ngc(k\dn\pm\nbin)}\right]_{\rgc}
\end{eqnarray}
\citep[Equations (28--30) and (32--34) in][]{leung2013}.

To complete the problem, we need only calculate the coefficients
$\Phi_k$ and $\Phi^e_k$ from Taylor expansions of the full
potential, $\Phi$. For the circumbinary case, the expansion variable
is $\abin/\rgc$, while for an \circpri\ orbit, we use
$\rgc/\abin$. For example, in the \circpri\ configuration, the
first few coefficients are
\begin{eqnarray}
\Phi_0 & = & -\frac{G\Mp}{\rgc} -
\frac{G\Ms}{\abin}\left(1+\frac{1}{4}\frac{\rgc^2}{\abin^2}
+\frac{9}{64}\frac{\rgc^4}{\abin^4}
+\frac{25}{256}\frac{\rgc^6}{\abin^6}+\dots\right),
\\
\Phi_1 & = &
-\frac{G\Ms}{\abin}\left(
\frac{3}{8}\frac{\rgc^3}{\abin^3}
+\frac{15}{64}\frac{\rgc^5}{\abin^5}+\dots\right),
\\
\Phi_2 & = &
-\frac{G\Ms}{\abin}\left(\frac{3}{4}\frac{\rgc^2}{\abin^2}
+\frac{5}{16}\frac{\rgc^4}{\abin^4}
+\frac{105}{512}\frac{\rgc^6}{\abin^6}+\dots\right),
\end{eqnarray}
and
\begin{eqnarray}
\Phi^e_0 & = &
\frac{G\Ms}{\abin}\left(1+\frac{3}{4}\frac{\rgc^2}{\abin^2}
+\frac{45}{64}\frac{\rgc^4}{\abin^4}
+\frac{175}{256}\frac{\rgc^6}{\abin^6}+\dots\right),
\\
\Phi^e_1 & = &
\frac{G\Ms}{\abin}\left(\frac{3}{2}\frac{\rgc^3}{\abin^3}
+\frac{45}{32}\frac{\rgc^5}{\abin^5}+\dots\right),
\\
\Phi^e_2 & = &
\frac{G\Ms}{\abin}\left(\frac{9}{4}\frac{\rgc^2}{\abin^2}
+\frac{25}{16}\frac{\rgc^4}{\abin^4}
+\frac{735}{512}\frac{\rgc^6}{\abin^6}+\dots\right),
\end{eqnarray}
while the angular frequencies (Equations (\ref{eq:ngc}) and 
(\ref{eq:kappas})) are
\begin{equation}
\label{eq:meanmotionuni}
\ngc = \nkep
\left(1-\frac{1}{4}\frac{\Mjup}{\Msun}\frac{\rgc^3}{\ajup^3}
-\dots\right) ,
\end{equation}
\begin{equation}
\label{eq:freeapsidalfrequni}
\kappae = \nkep
\left(1-\frac{\Mjup}{\Msun}\frac{\rgc^3}{\ajup^3}
-\dots\right),
\end{equation}
\begin{equation}
\label{eq:freenodalfrequni}
\kappai = \nkep
\left(1+\frac{\Mjup}{\Msun}\frac{\rgc^3}{\ajup^3}
+\dots\right),
\end{equation}
where $\nkep$ is the Keplerian frequency ($\nkep^2 = G\Msun/\rgc^3$).
\citet{bk2015tatooine} provide a similar list for circumbinary orbits.

For comparison with test particle orbits in binaries derived from secular
perturbation theory \citep[e.g.,][]{hepp1974, murray1999}, we
rewrite orbit solutions to separate out non-secular terms. For example,
\begin{eqnarray}\label{eq:rsolncompsecular}
R(t) & \approx & \rgc \left[1 - \efree \cos(\kappae t+\psi_e) -
\eforce \cos(\ngc t) - F(t,\rgc,)\right] 
\end{eqnarray}
where $F$ is a cosine series with the non-secular part, and
motion with a corresponding \textrm{forced} eccentricity of
\begin{equation}\label{eq:eforceC}
\eforce = \myC^-_1\ebin \approx \frac{5}{4}\frac{\rgc}{\abin} \ebin 
\end{equation} 
with a similar expression for $\phi(t)$, taking advantage of the
approximation that $\myD^-_1 = 2\myC^-_1$, which 
neglects terms of order $(\Ms/\Mp)(\rgc/\abin)^4$.

\section{Eccentricity evolution from secular perturbation 
  theory}\label{appx:secular}

In contrast to the instantaneous orbit solutions of Lee-Peale-Leung 
theory, secular perturbation theory identifies long-term trends in the 
orbital elements of a planetesimal. We specialize our analysis to the
solar system, where the starting point is the disturbing function 
\citep{murray1999},
\begin{equation}
{\cal R} = -\Phi -\frac{G\Mjup}{\ajup^2} R \cos(\Delta\phi) + \frac{G\Msun}{R} ~ .
\end{equation}
Here $\Phi$ is the potential of the planetesimal as measured in a
reference frame tied to the Sun. The second term accounts for the
acceleration of that reference frame stemming from Jupiter; the
last term is the potential in the absence of the
perturbations. Thus $\cal{R}$ is a measure of the effect of perturbers
on the planetesimal's otherwise Keplerian motion.

To derive the secular changes in the planetesimal's orbit, we
obtain the orbit-averaged value of $\cal{R}$, expressed using
the Keplerian semimajor axis $a$, eccentricity $e$ and argument
of perihelion $\varpi$ (here we neglect
inclination $i$). Following convention, we introduce the variables
\begin{equation}
h = e\sin(\varpi)
 \ \ \ \mbox{\rm{and}} \ \ \ \ 
k = e\cos(\varpi);
\end{equation}
We then write the orbit-averaged perturbing function to first order 
in the eccentricity of Jupiter, and second order in the planetesimal's
eccentricity,
\begin{equation}
\Ravg
\approx 
\nkep^2 a^2 
\left[\frac{1}{2} A (h^2 + k^2) + B (k\cos \varpi_J - h\sin \varpi_J) + C
\right]
\end{equation}
where $\varpi_J$ is Jupiter's argument of perihelion, $A$ is the
precession rate of the planetesimal, $B$ is an eccentricity driving term,
and $C$ is a constant.  In the absence of perturbers other than
Jupiter,
\begin{equation}
A \rightarrow \ompresJ 
\approx \frac{3}{4}\nkep\frac{\Mjup}{\Msun}\frac{a^3}{\ajup^3},
\end{equation}
and 
\begin{equation}\label{eq:B}
B \rightarrow -\frac{15}{16}\ejup\nkep\frac{\Mjup}{\Msun}\frac{a^4}{\ajup^4}.
\end{equation}
Other perturbers add contributions linearly to these terms. In an axisymmetric 
(i.e., non-eccentric) disk, only the precession factor changes. We then add 
the disk-induced precession rate,
\begin{eqnarray}\label{eq:omprediskgen}
\ompredisk & = & -\left[\frac{1}{2\nkep R^2}\frac{d}{dR}
  \left(R^2\frac{d\Phi_d}{dR}\right)\right]_{R=a} 
\end{eqnarray} 
where $\Phi_d$ is the disk potential.

The evolution of the orbital elements derives from
the time-independence of the orbit-averaged perturbing function:
\begin{equation}
\frac{d}{dt}\Ravg = 
\frac{dk}{dt}\frac{\partial}{\partial k}\Ravg +
\frac{dh}{dt}\frac{\partial}{\partial h}\Ravg
= 0;
\end{equation}
Thus,
\begin{eqnarray}
\frac{dh}{dt} = \frac{1}{\nkep a^2}\frac{\partial}{\partial k}\Ravg 
 = A k + B \cos\varpi_J 
\\
\frac{dk}{dt} = -\frac{1}{\nkep a^2} \frac{\partial}{\partial h}\Ravg 
= -A h + B \sin\varpi_J.
\end{eqnarray}
We can extract from these expressions the overall evolution of
the eccentricity and argument of perihelion,
\begin{eqnarray}
\frac{de}{dt} & = & B \sin(\varpi - \varpi_J)
\\
\frac{d\varpi}{dt} & = & A + \frac{B}{e} \cos(\varpi - \varpi_J)
\end{eqnarray}
\citep{naga2005}.

Solutions to these evolution equations for small bodies, with constant
eccentricity and apsidal alignment with Jupiter, yield the forced
eccentricity,
\begin{equation}
\eforce = -\frac{B}{A}.
\end{equation}
The value of $A$ can be positive or negative. If negative, then a
body's forced eccentric orbit is anti-aligned with Jupiter.
More generally, the evolution equation describes motion with a 
forced eccentricity that does not precess and a free eccentricity 
that precesses at a rate given by $A$.

\section{Numerical Simulations of Planet Formation}
\label{appx:numsim}

To study the growth and evolution of rocky planets in the inner solar 
system, we rely on \orchestra, a parallel \verb!C++/MPI! hybrid 
coagulation + \nbody\ code that tracks the accretion, fragmentation, 
and orbital evolution of solid particles ranging in size from a few 
microns to thousands of km \citep{kb2008, bk2011a, kb2016a}. The ensemble 
of codes within \orch\ includes a multi-annulus coagulation code for 
small particles, an \nbody\ code for large particles, and a radial 
diffusion code to follow the evolution of the gaseous disk. Other 
algorithms link the codes together, enabling each component to react 
to the evolution of other components.

In the coagulation code
\citep{kb2001,kb2002a,kb2004a,kb2008,kb2010,kb2012,kb2015a}, we divide 
a circumstellar disk with inner radius \ain\ and outer radius 
\aout\ into $N$ concentric annuli with width $\Delta a_i$ centered at 
$a_i$. Within each annulus, there are $M$ mass batches with characteristic 
mass $m_k$ and logarithmic spacing $\delta = m_{k+1} / m_k$. Batches 
contain $N_{ik}$ particles with total mass $M_{ik}$, average mass
$\bar{m}_{ik} = M_{ik} / N_{ik}$, horizontal velocity $h_{ik}$, and vertical velocity
$v_{ik}$.  For this suite of calculations, we ignore the interactions of solid 
material with the gaseous component of the disk. Once we specify an initial 
distribution of masses in each annulus, the numbers and velocities of each 
batch evolve through physical collisions and gravitational interactions 
with all other mass batches in the disk.

To specify collision rates, we adopt the particle-in-a-box algorithm. 
For a particle in annulus $i$ and mass batch $k$, the collision rate 
with other particles is 
$N_{jl} ~ \sigma ~ v ~ f_g ~ p / V$
where $N_{jl}$ is the number of particles in another batch,
$\sigma$ is the geometric cross-section,
$v$ is the relative velocity,
$f_g$ is the gravitational focusing factor,
$p$ is the probability that particles interact, and
$V$ is the volume occupied by the particles \citep{kb2002a}.
The relative velocity depends on $h_{ik}$ and $v_{ik}$ for each batch.
When relative velocities are large (small), the gravitational focusing 
factor is derived in the dispersion (shear) regime \citep{kb2004a,kb2012}.
For batches in the same annulus, $p$ = 1; otherwise, $p$
depends on the fraction of overlapping volumes occupied by the 
particles \citep[e.g.,][]{kb2002a}.

Collision outcomes depend on the ratio of the center-of-mass collision
energy $Q_c$ and the collision energy required to eject half of the mass
to infinity \qdstar. When two particles collide, the mass of the merged 
particle is
\begin{equation}
m = m_{ik} + m_{jl} - \mesc ~ ,
\label{eq: msum}
\end{equation}
where the mass of debris ejected in a collision is
\begin{equation}
\mesc = 0.5 ~ (m_{ik} + m_{jl}) \left ( \frac{Q_c}{Q_D^*} \right)^{b_d} ~ 
\label{eq: mej2}
\end{equation}
and $b_d$ is a constant of order unity. Here, we adopt $b_d$ = 1 and
set fragmentation parameters in the \qdstar\ relation to those 
appropriate for rocky material:
$Q_b \approx 3 \times 10^7$~erg~g$^{-1}$~cm$^{-\beta_b}$,
$\beta_b \approx -0.40$, $Q_g \approx$ 0.3~erg~g$^{-2}$~cm$^{3-\beta_g}$,
and $\beta_g \approx$ 1.35 for particles with mass density $\rho_p$ 
= 3~\gcmc\ \citep[see also][]{davis1985,hols1994,love1996,housen1999,
ryan1999,arakawa2002,giblin2004,burchell2005}.
Particles in the debris have a power-law cumulative size distribution,
$n(>m) \propto m^{1/6}$, where the largest particle in the debris has
\begin{equation}
\mmaxd = m_{L,0} ~ \left ( \frac{Q_c}{Q_D^*} \right)^{-b_L} ~ \mesc ~ ,
\label{eq: mlarge}
\end{equation}
$m_{L,0} \approx$ 0.01--0.5, and $b_L \approx$ 0--1.25
\citep{weth1993,kb2008,koba2010a,weid2010}.
We adopt $m_{L,0} =$ 0.2 and $b_L \approx$ 1.

Within the coagulation grid, $h_{ik}$ and $v_{ik}$ evolve due to collisional 
damping from inelastic collisions and gravitational interactions.  For 
inelastic and elastic collisions, we follow the statistical, Fokker-Planck 
approaches of \citet{oht1992} and \citet{oht2002}, which treat pairwise 
interactions (e.g., dynamical friction and viscous stirring) between all 
objects.  We also compute long-range stirring from distant oligarchs 
\citep{weiden1989}.

Calculations begin with an initial mass distribution between minimum radius 
\rmin\ and maximum radius \r0\ in each annulus. The initial surface density 
is $\Sigma = \Sigma_0 (a /{\rm 1~AU})^{-3/2}$; 
$\Sigma_0$ = 10~\gcms\ is roughly the surface density of solids in the 
Minimum Mass Solar Nebula. The initial velocities $h_0$ and $v_0$ are
related to the initial orbital $e$ and $i$:
$h_0 = 0.79 e_0 V_K$ and $v_0 = {\rm sin}~i~V_K / \sqrt{2} $,
where $V_K$ is the local circular velocity. We set $e_0$ and $i_0$ to
yield relative velocities smaller than the escape velocity of the largest
object in the annulus and gravitational focusing factors in the dispersion
regime.

As calculations proceed, the algorithm sets the time step based on the changes 
to particle numbers and velocities in all of the mass bins. The algorithm has
been tuned to match analytic solutions to the coagulation equation
\citep[e.g.,][]{kl1998,kb2015a} and numerical solutions of solid evolution 
derived by other investigators \citep[e.g.,][]{weth1993,weiden1997b}. Overall,
solutions conserve mass and energy to machine accuracy over $10^7 - 10^8$ 
time steps. In addition to \citet{kl1998} and \citet{kb2015a}, 
\citet{bk2006}, \citet{bk2011a} and \citet{kb2016a} describe tests of 
the multiannulus coagulation code.

\bibliography{planets}{}
\bibliographystyle{aasjournal}

\end{document}